\documentclass[%
reprint,
superscriptaddress,
showpacs,preprintnumbers,
amsmath,amssymb,
prb,
]{revtex4-2}

\usepackage{graphicx}
\usepackage{dcolumn}
\usepackage{xfrac}
\usepackage{bm}
\usepackage{color}
\usepackage[export]{adjustbox}
\usepackage{xr-hyper}
\usepackage{hyperref}
\usepackage{tabularx}

\usepackage{amsmath}
\usepackage{amsfonts}
\usepackage{amssymb}
\usepackage{tabularx}
\usepackage{chemformula}
\usepackage{soul,xcolor}
\setstcolor{red}

\newcommand{\TC}{$T_{\text C}$}
\newcommand{\TN}{$T_{\text N}$}
\newcommand{\ub}{$\mu_{\text B}$}

\begin{document}

\title{Topological Hall effect induced by chiral fluctuations in \ch{ErMn6Sn6}}

\author{Kyle~Fruhling$^{\ast\dagger}$}
\affiliation{Department of Physics, Boston College, Chestnut Hill, MA 02467, USA}
\thanks{These authors contributed equally to this work.\\$\dagger$ Corresponding author: fruhling@bc.edu}

\author{Alenna~Streeter$^\ast$}
\affiliation{Department of Physics, Boston College, Chestnut Hill, MA 02467, USA}

\author{Sougata~Mardanya}
\affiliation{Department of Physics and Astrophysics, Howard University, Washington, DC 20059, USA}

\author{Xiaoping~Wang}
\affiliation{Neutron Scattering Division, Oak Ridge National Laboratory, Oak Ridge, TN 37831, USA}

\author{Priya~Baral}
\affiliation{Laboratory for Neutron Scattering and Imaging (LNS), Paul Scherrer Institut, PSI, Villigen, CH-5232, Switzerland}

\author{Oksana~Zaharko}
\affiliation{Laboratory for Neutron Scattering and Imaging (LNS), Paul Scherrer Institut, PSI, Villigen, CH-5232, Switzerland}

\author{Igor~I.~Mazin}
\affiliation{Department of Physics and Astronomy and Quantum Science and Engineering Center, George Mason University, Fairfax, VA 22030, USA}

\author{Sugata~Chowdhury}
\affiliation{Department of Physics and Astrophysics, Howard University, Washington, DC 20059, USA}

\author{William~D.~Ratcliff}
\affiliation{NIST Center for Neutron Research, National Institute of Standards and Technology, Gaithersburg, MD 20899-6100, USA}
\affiliation{Department of Physics and Department of Materials Science and Engineering, University of Maryland, College Park, MD 20742, USA}

\author{Fazel~Tafti}
\affiliation{Department of Physics, Boston College, Chestnut Hill, MA 02467, USA}


\begin{abstract}
Topological Hall effect (THE) is a hallmark of scalar spin chirality, which is found in static skyrmion lattices.
Recent theoretical works have shown that scalar spin chirality could also emerge dynamically from thermal spin fluctuations.
Evidence of such a mechanism was found in the kagome magnet \ch{YMn6Sn6} where fluctuations arise from frustrated exchange interactions between Mn kagome layers. 
In \ch{YMn6Sn6}, the rare-earth ion Y$^{3+}$ is non-magnetic.
When it is replaced by a magnetic ion (Gd$^{3+}$--Ho$^{3+}$), the intrinsically antiferromagnetic Mn-Mn interlayer coupling is overwhelmed by the indirect ferromagnetic Mn-$R$-Mn one, relieving frustration. 
This generates interesting anomalous Hall conductivity, but not THE. 
Here we show that Er lies in an intermediate regime where direct and indirect interactions closely compete, so that \ch{ErMn6Sn6} can switch from one regime to the other by temperature, i.e., from a collinear ferrimagnetic ground state to a spiral antiferromagnet at 78~K.
The AFM phase forms a dome in the temperature-field phase diagram.
Close to the boundary of this dome, we find a sizable fluctuations-driven THE, thus underscoring the universality of this chiral fluctuation mechanism for generating non-zero scalar spin chirality.
\end{abstract}

\maketitle


\section{\label{sec:introduction}Introduction}
The unique geometry of the 2D kagome lattice leads to appreciable frustration in the nearest neighbor magnetic or electronic tight-binding models.
While numerous materials with antiferromagnetic (AFM) kagome planes, usually correlated insulators, have been investigated for spin-liquid behavior~\cite{norman_colloquium_2016}, another class of interest is stacked ferromagnetic kagome layers, which retains the peculiarity of the original kagome model in terms of electronic structure, but not magnetic frustration~\cite{Ghimire2020}.
The electronic structure of such materials is described by a frustrated hopping model, which was originally restricted to $s$-orbitals only~\cite{tang_high-temperature_2011}, but later extended to $d$-orbitals in compounds with hexagonal symmetry~\cite{lee_interplay_2023}. 
Since many of these materials are magnetically ordered above room temperature and easily manipulated by doping, they provide a fascinating playground for topology and magnetism to interact on a kagome lattice and produce exotic behaviors in momentum space, such as flat bands and Dirac crossings~\cite{mazin_theoretical_2014,bolens_topological_2019,kang_dirac_2020}.

One such family of compounds, dubbed ``166'', has attracted particular attention. 
The general formula is $RT_6M_6$, where $R$ is a rare-earth, $T$ is a 3d transition-metal, usually Mn, and $M$ is a metalloid of the group 13 or 14, most commonly Sn. 
In all these compounds, the Mn layers are metallic and inherently ferromagnetic (FM); however, the coupling between layers and the anisotropy of the ordered state depend on the rare-earth atom between the layers~\cite{venturini_magnetic_1991,el_idrissi_magnetic_1991,gorbunov_magnetic_2012}.
For example, \ch{TbMn6Sn6} has a strong Tb-Mn coupling and shows an out-of-plane ferrimagnetic (FIM) order, which leads to a sizable anomalous Hall effect (AHE)~\cite{
xu_topological_2022,riberolles_low-temperature_2022,mielke_iii_low-temperature_2022,jones_origin_2022}. 
The non-magnetic rare-earths (Y and Sc) have no $R$-Mn coupling and favor a spiral antiferromagnetic (AFM) coupling between Mn layers, which leads to a sizable topological Hall effect (THE)~\cite{dally_chiral_2021,ghimire_competing_2020,zhang_magnetic_2022}.
In this paper, we discuss the peculiar case of \ch{ErMn6Sn6}, a relatively less studied member of the 166 family~\cite{casey_spin-flop_2023}, which falls in a regime between the FIM \ch{TbMn6Sn6} and spiral AFM \ch{YMn6Sn6}.

As we will show here, the peculiarity of \ch{ErMn6Sn6} lies in the strength of the Er-Mn coupling, which is non-zero unlike Y-Mn, but much weaker than Tb-Mn, so the net sign of the Mn-Mn coupling can be reversibly switched by reducing the ordered Er moment through thermal fluctuations.
This leads to a change of magnetic state from FIM at low temperatures (similar to \ch{TbMn6Sn6}) to spiral AFM at high temperatures (similar to \ch{YMn6Sn6}).
Using field-dependent magnetization and neutron diffraction data, we show that the spiral AFM state occupies a dome in the $H$-$T$ phase diagram.
We also show that the field-induced transition between these two states involves an intermediate regime of fluctuating Mn moments.   
By performing detailed magneto-transport measurements, we reveal a sizable THE in this fluctuating regime, i.e., at the boundary between the spiral AFM phase (where \ch{ErMn6Sn6} is analogous to \ch{YMn6Sn6}) and FIM phase (where it is analogous to \ch{TbMn6Sn6}).
The magnitude of THE increases with increasing temperature as expected from the fluctuation-driven mechanism.

\section{\label{sec:results}Results}
Details of the crystal growth, magnetization and transport characterizations, neutron scattering, and DFT calculations are explained in the Supplemental Material~\cite{suppmatt} with appropriate references~\cite{rodriguez-carvajal_recent_1993, momma_vesta_2011, hohenberg1964inhomogeneous, kresse1996efficient, kresse1999ultrasoft, perdew1996generalized, Monkhorst1976, allen2014occu, coates_suite-level_2018, zikovsky_crystalplan_2011, schultz_integration_2014, Petricek_Jana_2023}.

\subsection{\label{subsec:Magnetization}Magnetization} 
\begin{figure*}
  \includegraphics[width=\textwidth]{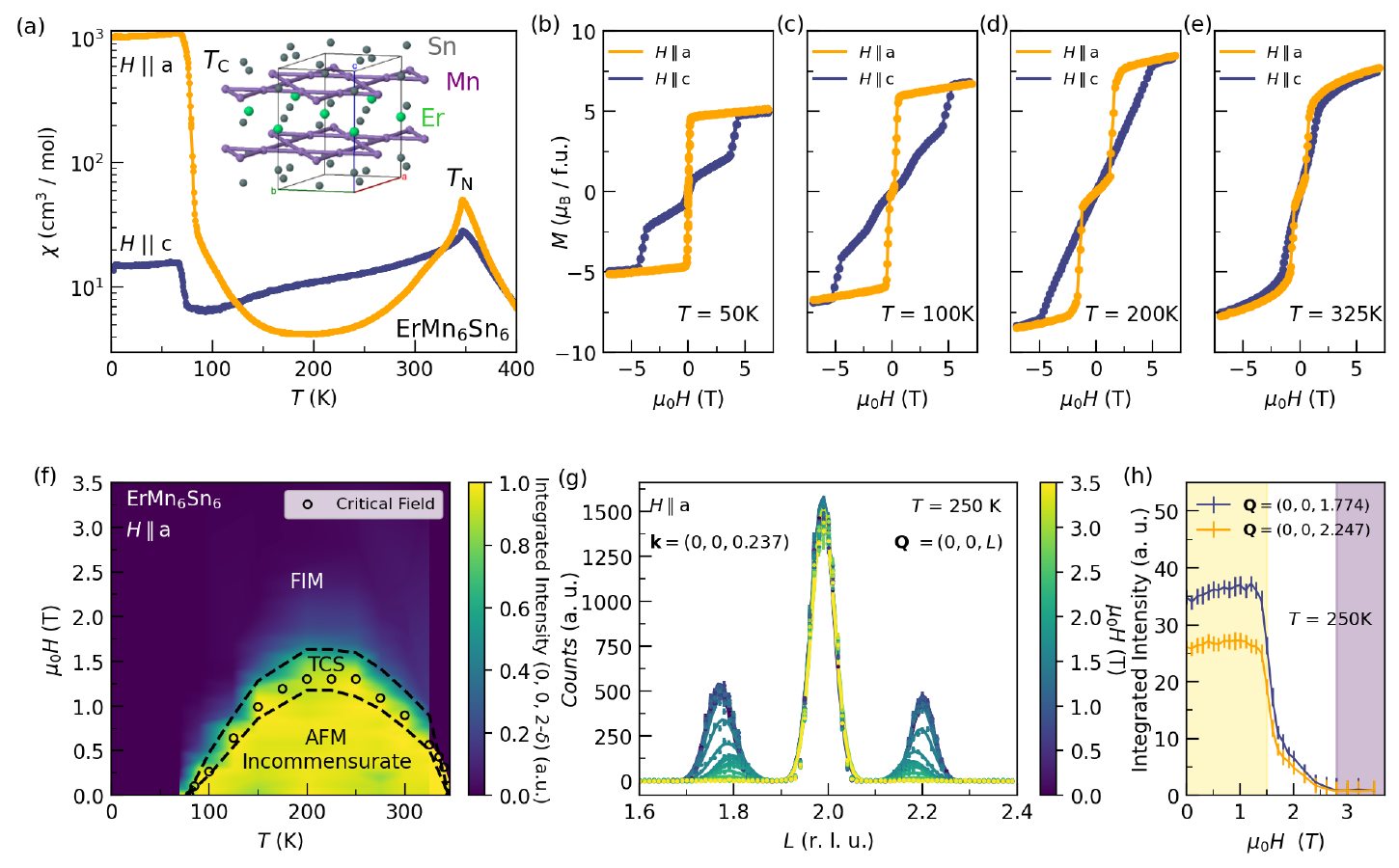}
  \caption{\label{fig:MAG}
 (a) Magnetic susceptibility of \ch{ErMn6Sn6} as a function of temperature measured in both in-plane and out-of-plane field directions and plotted on a semilog plot. Zero-field cooled susceptibility measurements taken with $H=100$~Oe. Inset: Crystal structure of \ch{ErMn6Sn6} with Sn atoms in silver Mn atoms in purple and Er atoms in green.
 (b-e) Magnetization as a function of field at several temperatures.
 (f) Magnetic phase diagram constructed from the in-plane magnetization data (black circles) and neutron diffraction data (color map). The color map represents normalized integrated intensity under incommensurate AFM Bragg peaks.
 (g) Neutron diffraction $\mathbf{Q}$ scan in the (00L) direction showing a structural Bragg peak at the center and two satellite magnetic peaks. The satellites indicate incommensurate AFM order at $\mathbf{Q}=(0,0,2\pm\mathbf{k})$ with $\mathbf{k}=0.237$. 
 (h) Field dependence of the integrated intensity under the satellite peaks in panel g reveal three regimes at 250~K. Error bars in this paper represent one standard deviation. 
  }
\end{figure*}
The crystal structure of \ch{ErMn6Sn6} is shown in the inset of Fig.~\ref{fig:MAG}a. We determined the crystal structure at 40 K and 200 K from X-ray diffraction and found it to be hexagonal (space group $P6/mmm$) with a- and c-axis parameters being 5.49 and 8.97~\AA\ (details are presented in Table S1). The Mn atoms (purple spheres in Fig.~\ref{fig:MAG}a) form
a kagome network, whereas the Er atoms (light green spheres) form a triangular lattice. The Sn atoms (grey spheres) are dispersed between the Er and Mn sublattices.

\ch{ErMn6Sn6} has a high-temperature AFM transition marked by a peak at \TN$=$346(3)~K in Fig.~\ref{fig:MAG}a and a low-temperature FIM order marked by a step in the in-plane susceptibility at \TC$=$78(8)~K.
The AFM ordering has less anisotropy than the FIM order which has a marked easy-plane character, as seen in Fig.~\ref{fig:MAG}a.
The evolution of anisotropy with temperature is demonstrated in the magnetization curves of Figs.~\ref{fig:MAG}b-e.
At low temperatures, the out-of-plane magnetization ($H\|c$) saturates at a much higher field than in-plane magnetization ($H\|a$).
However, this anisotropy becomes less pronounced as temperature is increased from \TC$=$78~K to \TN$=$346~K. 

Using the magnetization curves in Figs.~\ref{fig:MAG}b-e and the supplementary Fig.~S1, we construct the magnetic phase diagram of \ch{ErMn6Sn6} for $H\|a$ and $H\|c$, respectively, in Fig.~\ref{fig:MAG}f and the supplementary Fig.~S2~\cite{suppmatt}.
Focusing on the orange curves ($H\|a$) in Figs.~\ref{fig:MAG}b-e, the $M(H)$ curve at 50~K (Fig.~\ref{fig:MAG}b) shows a rapid saturation of the magnetic moment characteristic of FIM ordering.
However, the $M(H)$ curves at higher temperatures (Figs.~\ref{fig:MAG}c-e) show a linear regime at low fields characteristic of AFM ordering, followed by a saturation due to a field-induced transition from the AFM to FIM order.
The black data points in Fig.~\ref{fig:MAG}f trace this field-induced transition between AFM (linear regime) and FIM (saturated regime) orders, revealing an AFM dome in the phase diagram.

\subsection{\label{subsec:Neutron}Neutron Diffraction}
\begin{figure}
  \includegraphics[width=\columnwidth]{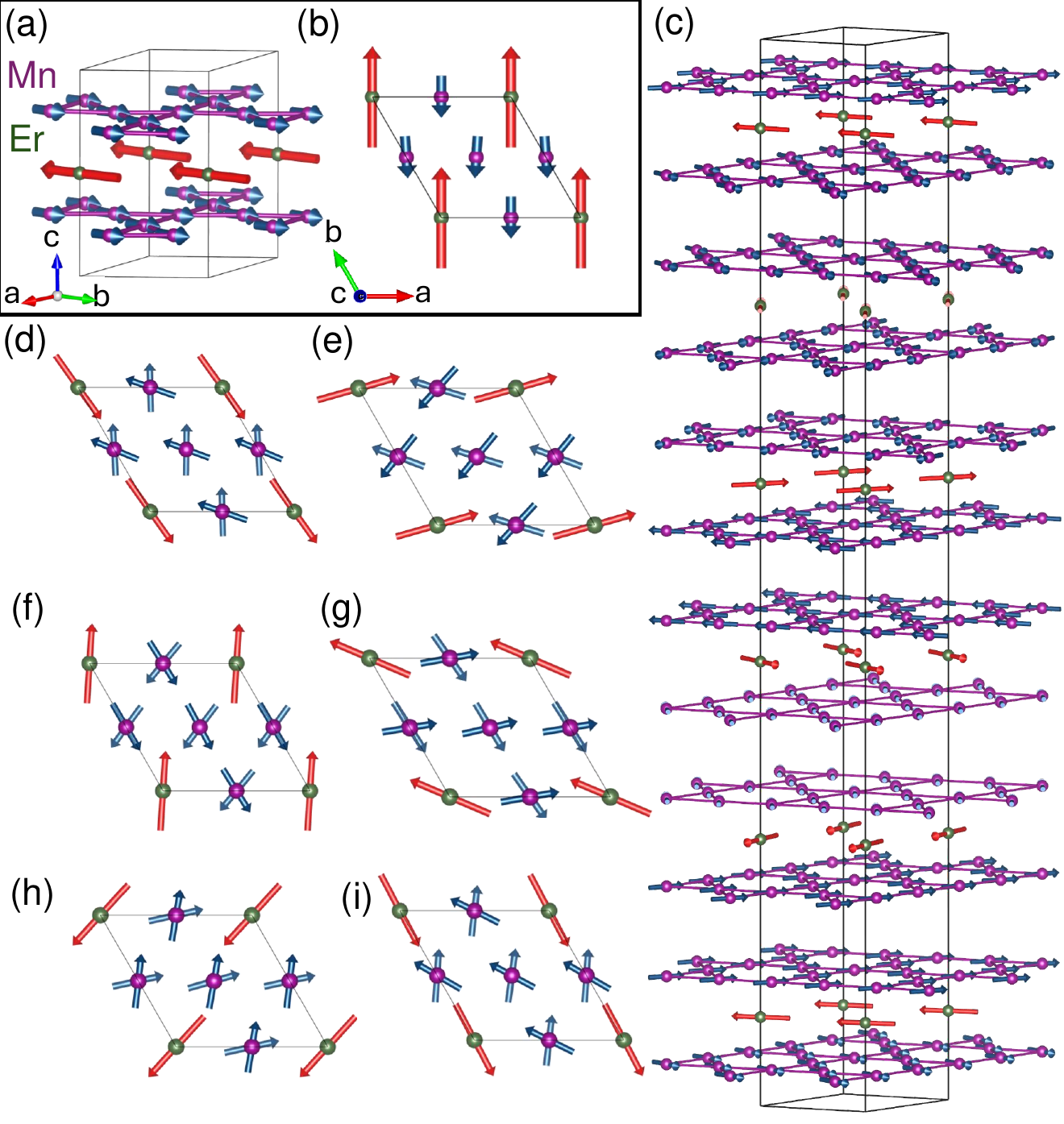}
  \caption{\label{fig:STRUCT}
  Magnetic moments in \ch{ErMn6Sn6}. Nonmagnetic Sn atoms are removed for clarity.
 (a) Collinear FIM phase unit cell moments.
 (b) FIM phase unit cell looking down the c-axis.
 (c) Six layer super-cell of the incommensurate spiral AFM phase.
 (d-i) Individual layers of the AFM phase looking down the c-axis showing the evolution of the orientation of the Er and Mn moments. Note that the orientation does not quite repeat after five layers as this phase is incommensurate.
  }
\end{figure}
We used single crystal neutron diffraction to characterize the FIM and AFM structures, and study their evolution with magnetic field in the phase diagram of Fig.~\ref{fig:MAG}f.
It is instructive to start with the AFM dome at the center of the phase diagram.
Fig.~\ref{fig:MAG}g shows diffraction patterns at several fields along the $\mathbf{Q}=(0,0,L)$ direction at 250~K.
The central peak at $\mathbf{Q}=(0,0,2)$ is a structural Bragg peak and the two satellites at $\mathbf{Q}=(0,0,2\pm\mathbf{k})$, with $\mathbf{k}=0.237$ at zero-field, are incommensurate AFM Bragg peaks.
Previous diffraction studies at zero field~\cite{malaman_magnetic_1999} show that upon decreasing temperature, these satellite peaks move closer to the structural $(0,0,2)$ peak and merge with it at \TC$=$78~K, the temperature of the AFM to FIM transition at $H=0$. 

We studied the evolution of satellite peaks not only with temperature but also with in-plane magnetic field ($H\|a$).
As seen in Fig.~\ref{fig:MAG}g, the incommensurate satellite peaks are suppressed with increasing in-plane field and vanished at 3~T.
By tracing the integrated intensity under these satellite peaks as a function of field at 250~K, we reveal three regimes in Fig.~\ref{fig:MAG}h shaded with orange, white, and purple colors.
In the low-field region ($H<1.5$~T, orange), we find a nearly unchanged magnetic intensity under the satellites peaks in the incommensurate AFM phase.
In the high-field region ($H>2.8$~T, purple), the satellite peaks are completely suppressed and the system is in the field-induced FIM phase.
In the intermediate region ($1.5<H<2.8$~T, white), the satellite peaks are gradually suppressed.
By repeating this analysis at different temperatures (supplementary Figs.~S3 and S4), we constructed the phase diagram of Fig.~\ref{fig:MAG}f where the color scale corresponds to the integrated intensity under the satellite peaks at different temperatures.
The yellow region is the incommensurate AFM phase where the satellite peaks are sizable.
The purple region is the FIM phase where satellite peaks have vanished.
The green region is the intermediate phase where the satellite peaks are being suppressed with field.

A refinement of the neutron diffraction data in the FIM phase yields the collinear in-plane magnetic structure shown in Fig.~\ref{fig:STRUCT}a,b.
Details of the magnetic refinements are presented in the supplemental Figs.~S5 and S6 and Tables~S1 to S4.
The Er and Mn magnetic moments are oppositely oriented with the respective magnitudes of 8.0~\ub\ and 2.3~\ub.
This structure is consistent with the zero-field ground state of \ch{ErMn6Sn6} reported in a prior work~\cite{malaman_magnetic_1999}.

The refinement analysis in the AFM phase at 200 K with an incommensurate $k$-vector (0, 0, 0.1959) produces a spiral order made of Mn-Er-Mn sandwich layers which rotate by 70.4$^\circ$ with respect to each other. The magnetic unit cell is approximately five times the structural unit cell (Fig.~\ref{fig:STRUCT}c-i). 
Within a single Mn-Er-Mn sandwich layer, the Mn moments from the top and bottom Mn layers form an angle $\phi$ and their net magnetization is canceled by the oppositely oriented Er moments.
This is the same double spiral order which has been found in \ch{YMn6Sn6}, now with a moment on the rare-earth site in \ch{ErMn6Sn6}~\cite{dally_chiral_2021}.
As seen in Fig.~\ref{fig:MAG}h, the transition between the spiral AFM phase and FIM phase is not abrupt; it involves an intermediate regime of what we believe to be fluctuating Mn moments. 
In the phase diagram of Fig.~\ref{fig:MAG}f, this fluctuating regime appears as a green region between the yellow (spiral AFM) and blue (FIM) regions.
From the available diffraction data one cannot give any evidence for the fluctuating behavior of the intermediate state, but we can speculate about it based on the observation of a sizable THE in the next section. Such a THE is prohibited in the AFM and FIM phases by symmetry but allowed in a transverse conical spiral (TCS) phase.

\subsection{\label{subsec:Topological}Topological Hall Effect} 
\begin{figure}
  \includegraphics[width=\columnwidth]{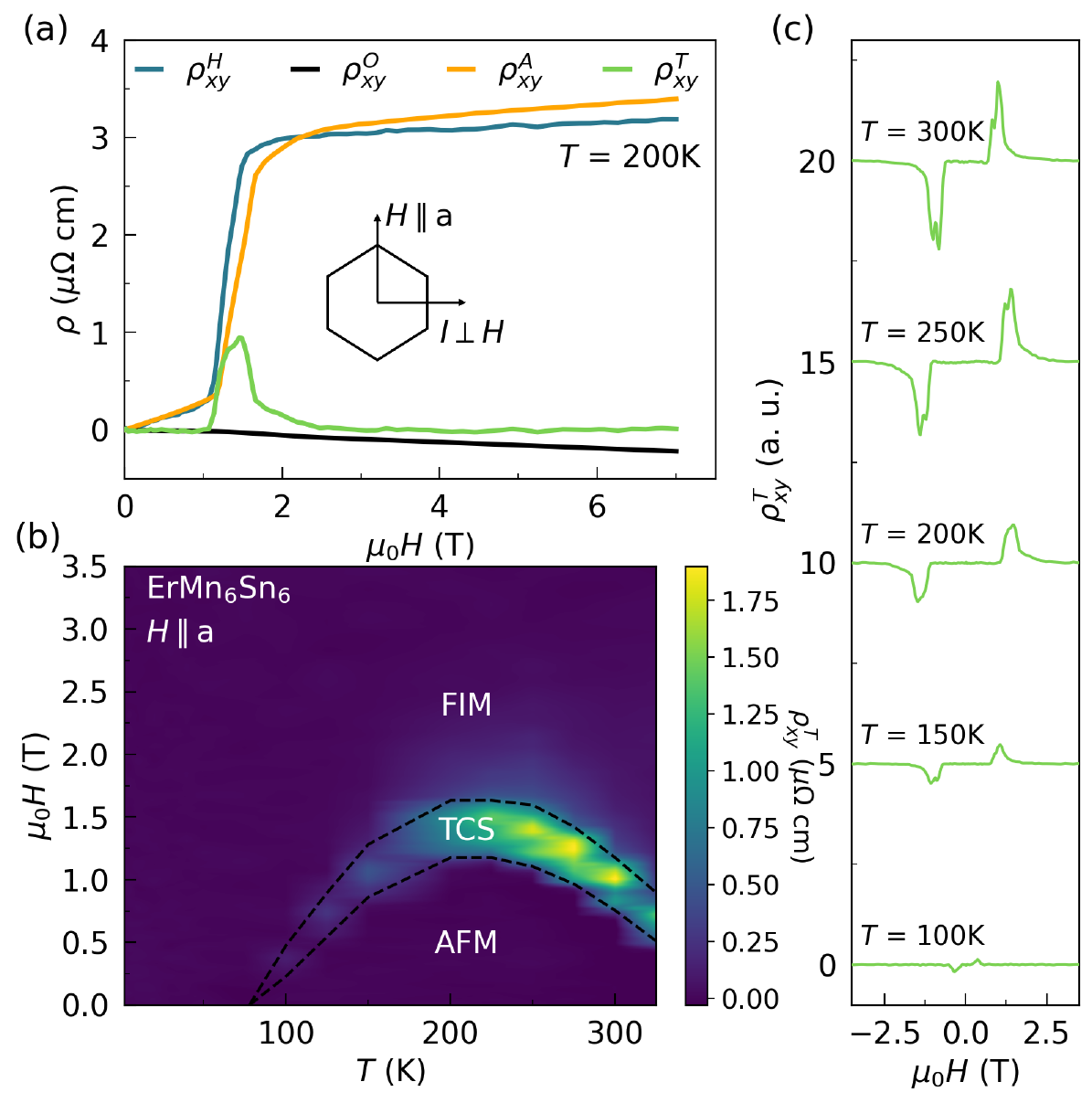}
  \caption{\label{fig:THE}
 (a) Topological Hall resistivity ($\rho_{xy}^T$) is extracted by subtracting the ordinary ($\rho_{xy}^O$) and anomalous ($\rho_{xy}^A$) Hall resistivities from the total Hall resistivity ($\rho_{xy}^H$). Inset: Diagram of field and current directions applied to the crystal.
 (b) A second version of the phase diagram (Fig.~\ref{fig:MAG}f) with a color map that corresponds to the intensity of the THE.
 (c) Representative $\rho_{xy}^T(H)$ curves at different temperatures.
  }
\end{figure}
The known mechanism of THE is due to the static scalar spin chirality (SSC) in non-coplanar magnetic structures, for example, in a skyrmion lattice~\cite{kurumaji_skyrmion_2019,shao_topological_2019,verma_unified_2022}. 
Recently, a dynamical mechanism for the THE has been proposed where spiral fluctuations, complementing an existing static spiral, generate a finite SSC and thus a THE~\cite{ghimire_competing_2020,wang_field-induced_2021,Fe3Ga}.
In the 166 compounds, this mechanism is operative only at finite temperatures, and only in a magnetic phase that is a transverse conical spiral (TCS) propagating along the $c$-axis. In a collinear AFM system the application of a magnetic field in the direction of sublattice magnetization leads to a spin-flop transition. When the AFM system has a spiral structure, it follows similarly that a cycloidal spin-flop, which is the TCS structure, occurs when an external magnetic field, $\mathbf{H}$, is applied in the $ab$-plane, a condition that is realized at the AFM phase boundary in \ch{ErMn6Sn6} (Fig.~\ref{fig:MAG}f).
THE is proportional to the emerging scalar spin chirality effective field, defined as 
$b_x=\mathbf{M}\cdot\nabla_y\mathbf{M}\times\nabla_z\mathbf{M}$. A single spin spiral propagating along $z$ cannot have a nonzero \textbf{b}, since $\mathbf{M}$ does not vary in the $xy$ plane. However, as pointed out in~\cite{ghimire_competing_2020}, a combination of a TCS propagating along $z$ and a flat spiral propagating along $y$ does generate an emerging field $b_x$. Since this field couples with an external magnetic field $H_x$, at finite temperature and in a fixed in-plane external field, thermally excited spiral magnons propagating in one direction, say, $y$ will have a preference over those propagating in the opposite direction, $-y$.
A simple thermodynamic calculation~\cite{ghimire_competing_2020} shows that such thermal excitations give a THE that is proportional to the applied field, $H$, temperature, $T$, and $M_c^2=M_{s}^2-M^2$, where $M_s$ is the saturated magnetization and $M_c$ is the projection of the Mn moment onto the $c$ direction, 
\begin{equation}
\label{eq:THE}
\rho^{T}=\kappa\, M_c^2TH 
\end{equation}
where $\rho^T$ is the topological hall resistivity and $\kappa$ is a constant of proportionality. In a prior study of \ch{YMn6Sn6} with non-magnetic Y$^{3+}$ ions~\cite{ghimire_competing_2020}, the variable $M_c$ was readily available because Mn was the only magnetic species, and the measured saturation magnetization was a good estimate of $M_s$, while the field-dependent magnetization was $M$. 
Then, the prefactor $\kappa$ in Eq.~\ref{eq:THE} would be a constant. 
In \ch{ErMn6Sn6}, however, Er$^{3+}$ ions carry a large and strongly temperature-dependent magnetization which prevents us from using Eq.~\ref{eq:THE} directly, even though it can be used separately at each individual temperature.
In other words, $\kappa$ becomes temperature dependent in \ch{ErMn6Sn6}.
For simplicity, we first present a more conventional method of extracting the THE and leave the implications of Eq.~\ref{eq:THE} for the Discussion.

We extract the THE from the total Hall signal as shown in Fig.~\ref{fig:THE}a, by subtracting the ordinary ($\rho^O_{xy}\propto H$) and anomalous ($\rho^A_{xy}\propto M$) Hall contributions to the transverse resistivity from the total signal ($\rho_{xy}^H$) to arrive at the topological Hall effect ($\rho^T_{xy}$) assuming $\rho^H=R_0H+4\pi R_sM+\rho^T$. 
In the high field FIM state, this assumption holds well and this equation simplifies to $\rho^H=R_0H+4\pi R_sM$.
Then following the procedure in Ref.~\cite{ghimire_competing_2020}, we can determine the high field anomalous and ordinary Hall effects by plotting $\rho^H/M$ as a function of $H/M$ and extracting the slope $R_0$ and intercept $4\pi R_s$. 
The typical treatment would assume these are both constant; however, the electronic structure and hence $R_0$ may be different in different magnetic phases. 
It also assumes that the anomalous Hall effect remains proportional to the magnetization which is not a fair assumption for spiral magnets.
In the treatment we follow we determine the ordinary Hall effect at low field as $\rho^O=\rho^H-4\pi R_sM$, and then interpolate a cubic spline for $\rho^O$ between the AFM and FIM states. 
The sum of the anomalous and ordinary Hall components is then assumed to vary smoothly between phases. Thus the sum can be used to determine $\rho^T$ even if the individual components are not accurately separated in the AFM state.
$\rho^T$ is then given by $\rho^T=\rho^H-R_0H-4\pi R_sM$.
The black, blue, and green lines in Fig.~\ref{fig:THE}a show the ordinary, anomalous, and topological components of the total Hall resistivity.
We repeated this analysis at different temperatures and made a color map of THE ($\rho^T$) in Fig.~\ref{fig:THE}b.
Our Hall data at different temperatures are presented in the supplementary Fig.~S7.
Since THE can be artificially generated due to the demagnetization effect, we measured both the Hall resistivity and magnetization data on the same sample.

\subsection{\label{subsec:Calculations}First-Pinciples Calculations} 
\begin{figure*}
  \includegraphics[width=\textwidth]{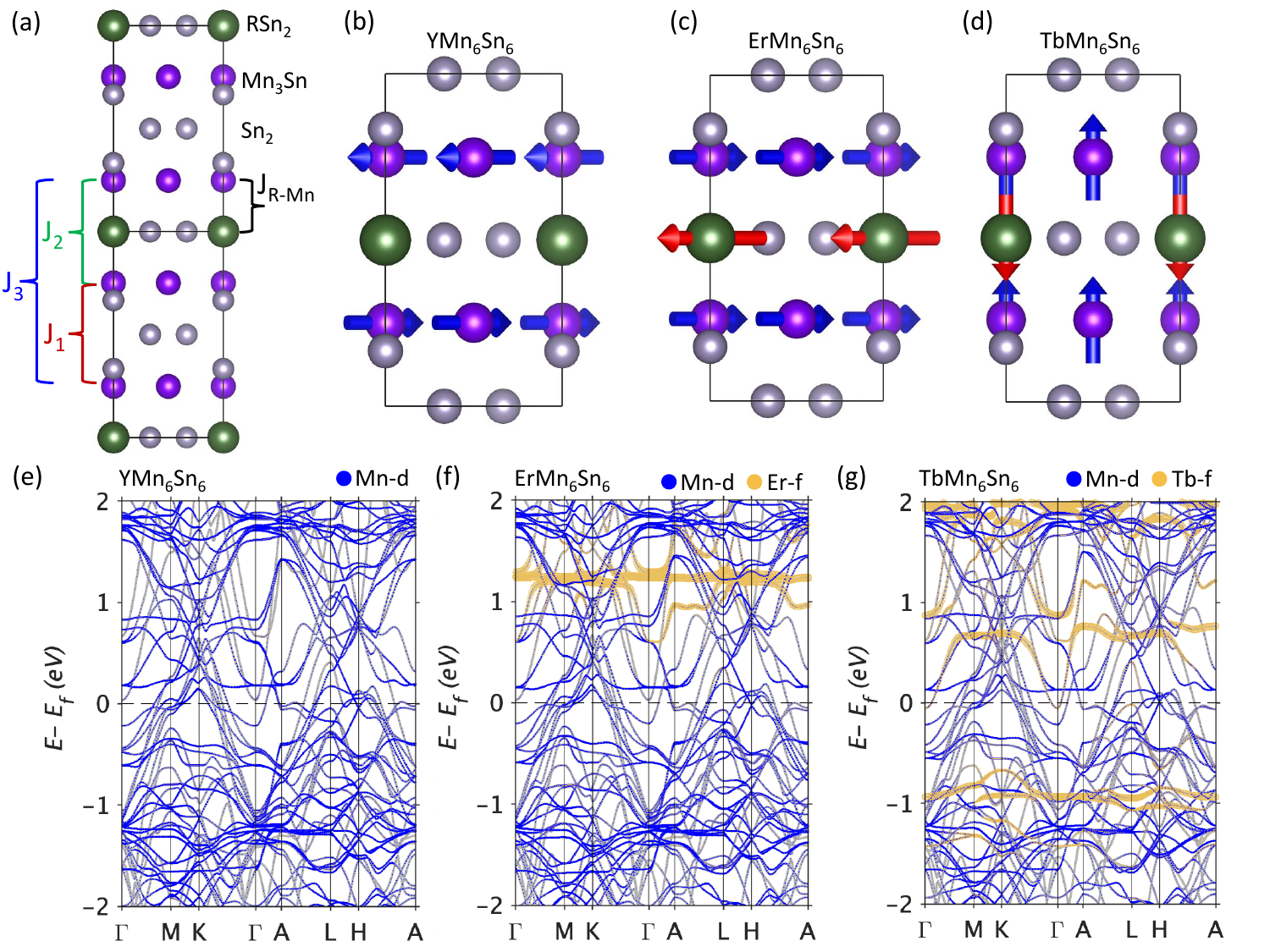}
  \caption{\label{fig:DFT}
 (a) Magnetic interactions in the $R$\ch{Mn6Sn6} system. 
 The ground state magnetic configuration for (b) \ch{YMn6Sn6}, (c) \ch{ErMn6Sn6}, and (d) \ch{TbMn6Sn6}. 
 (e-g) The orbital projected electronic structure with spin-orbit coupling along the high symmetry path of the Brillouin zone in \ch{YMn6Sn6}, \ch{ErMn6Sn6}, and \ch{TbMn6Sn6}. 
 The rare-earth f-states are highlighted in orange.
 Er $f$-bands at energies less than $-2$~eV exist but are invisible at this scale.
  }
\end{figure*}
To understand the TCS magnetic phase and the emerging THE in \ch{ErMn6Sn6}, we compare it to its sister compounds \ch{YMn6Sn6} and \ch{TbMn6Sn6}.
All three compounds share the same configuration of \ch{Mn3Sn} Kagome layers (or simply the Mn layers) sandwiched between two inequivalent \ch{Sn2} and $R$Sn$_2$ ($R$= Rare-earth) layers as shown in Fig.~\ref{fig:DFT}a. 
Isolated Mn kagome layers have a tendency for FM ordering.
Previous studies \cite{jones_origin_2022,lee_interplay_2023} have shown that the competition between the magnetic anisotropy of the rare-earth elements and Mn dictates the magnetic phase of these compounds. 
\ch{YMn_6Sn_6} and \ch{TbMn_6Sn_6} lie at two extreme ends of this picture.
In this case, the spiral ground state is understood by an effective model~\cite{ghimire_competing_2020} with three interlayer couplings, namely a ferromagnetic $J_1$ (across the \ch{Sn2} layer), antiferromagnetic $J_2$ (across the $R$Sn$_2$ layer), and a ferromagnetic $J_3$ ({second-neighbor interlayer coupling in} Fig.~\ref{fig:DFT}a).
$J_{R-\ch{Mn}}$, the coupling between \ch{Mn} and the rare-earth, can integrated out of the model, effectively providing a strongly rare-earth dependent FM contribution to the otherwise AFM $J_2$. In \ch{YMn6Sn6}, the spins are all aligned in the ab-plane, so the system has an easy plane.
Replacing Y with the magnetic atom Tb changes the magnetic anisotropy of the system into a strong easy axis. 
Therefore, the strong coupling from magnetic rare-earth enforces a FIM collinear magnetic state with out-of-plane moments for \ch{TbMn6Sn6}.

\ch{ErMn_6Sn_6} lies in an intermediate regime between the Y and Tb compounds. 
A comparison between the electronic structure of the three compounds is presented in Figs.~\ref{fig:DFT}e-g. 
In these calculations, the strong correlation effects of the $f$ and $d$ electrons are included by considering an effective onsite Hubbard potential ($U_{\rm eff} = U -J $ )~\cite{hubbard_U, Anisimov_1997}. 
We use $U^{\textrm{Er}}_{\rm eff}$ = 8~eV, $U^{\textrm{Tb}}_{\rm eff}$ = 7~eV, following the recommendation of Ref.~\cite{Galler_2022}, and $U^{\textrm{Mn}}_{\rm eff}$ = 0.5~eV. 
Close to the Fermi level ($E_\textrm{F}$), the electronic structures of the Y and Er compounds look nearly identical. 
This is because the Er-$f$ states lie well above the $E_\textrm{F}$ (1.3 eV) and do not have much influence on the bands near the Fermi level. 
In contrast, the Tb-$f$ states come closer to the $E_\textrm{F}$ and enhance the Tb-Mn interaction.
Therefore, the $R$-Mn exchange coupling is expected to be smaller in \ch{ErMn6Sn6} than in \ch{TbMn6Sn6}.  

To verify this conjecture, we calculated the exchange coupling constants and found that $J_\textrm{Tb-Mn}$ is at least four times larger than $J_\textrm{Er-Mn}$ (140~meV vs. 32~meV). 
Given that the direct Mn-Mn exchange coupling $J_2$ is AFM (and only weakly dependent on the rare earth) but the indirect coupling is FM (Fig.~\ref{fig:DFT}a), the net interaction appears antiferro- or ferromagnetic, and the ground state, correspondingly, spiral or ferrimagnetic, depending on which component dominates.
In the case of \ch{ErMn6Sn6}, at zero temperature, the FM component dominates, and the ground state is the same as in \ch{TbMn6Sn6}.
However, at higher temperatures, the Er moment starts to fluctuate (much more so than Mn, as shown in Ref. \cite{jones_origin_2022}), so the indirect exchange coupling $J_\textrm{R-Mn}$ is reduced compared to the direct coupling $J_2$, and eventually, the system switches to a spiral state, similar to
\ch{YMn6Sn6}.
With increasing field, this helical spiral transitions to a flopped TCS state, as suggested by our neutron diffraction and transverse resistivity data (the intermediate regime in Fig.~\ref{fig:MAG}h highlighted in white).
Note that, unlike
\ch{TbMn6Sn6}, \ch{ErMn6Sn6} is, both experimentally and computationally~\cite{lee_interplay_2023}, easy-plane, so no spin-reorientation transition happens in the Er system, unlike the Tb system~\cite{jones_origin_2022}.

\section{\label{sec:discuss}Discussion}
By mapping the temperature and field dependence of the THE in \ch{ErMn6Sn6} (Fig.~\ref{fig:THE}b), we reveal two characteristics that point toward a fluctuating order as the underlying mechanism of the THE.
First, the THE is observed only at the boundary between the spiral and ferrimagnetic phases.
Second, its magnitude increases with increasing temperature as expected from Eq.~\ref{eq:THE}.
In the supplementary materials, we use Eq.~\ref{eq:THE} to extract the THE signal and reproduce the phase diagram of Fig.~\ref{fig:THE}c (Figs.~S12 and S13), confirming its connection with Mn spin fluctuations.

The field dependence of the magnetic satellite peak intensities in Figs.~\ref{fig:MAG}f,h show that an intermediate fluctuating magnetic phase exists between the spiral and ferrimagnetic phases. From the temperature evolution and location of the THE at the AFM-FIM boundary, this phase appears to be a TCS. This is further supported by the expected~\cite{ghimire_competing_2020} drop in neutron intensity by a factor of slightly more than 2 in the intermediate phase Fig.~\ref{fig:MAG}h.
The THE emerges from the fluctuations of Mn spins in this phase.
Interestingly, the underlying spiral AFM state, the dome in Fig.~\ref{eq:THE}d, is in and of itself a result of fluctuations, but this time Er instead of Mn. 
A prerequisite for that is reduced Er-Mn coupling, compared to Tb-Mn.
As seen in Fig.~\ref{fig:DFT}, the Tb-f bands are too close to $E_\textrm{F}$ in \ch{TbMn6Sn6} unlike the Er-f bands in \ch{ErMn6Sn6}, confirming the above picture.
Thus, both the nontrivial magnetic phase diagram and the THE are driven by spin fluctuations, albeit of Er moments in the former, and Mn moments in the latter.

Note that thermal fluctuations are present in the AFM, TCS, and FIM phases (all three phases in the phase diagram of Fig.~\ref{fig:THE}b); however, they cannot produce a THE in either AFM or FIM phases since it is prohibited by symmetry. Only in the TCS, can thermal fluctuations  generate scalar spin chirality and thus a topological Hall effect.

\section*{ACKNOWLEDGMENTS}
The authors thank M.~Newburger and M.~Page for insightful discussions, and O.~Remcho, X.~Yao, and T.~Hogan for assistance with the experiments. 
The work at Boston College was supported by the National Science Foundation under Award No. DMR-2203512. 
The work at Howard University was supported by the U.S. Department of Energy (DOE), Office of Science, Basic Energy Sciences under Award No. DE-SC0022216.  
This research at Howard University used resources of the National Energy Research Scientific Computing Center, a DOE Office of Science User Facility supported by the Office of Science of the U.S. Department of Energy under Contract No. DE-AC02-05CH11231 using NERSC award BES-ERCAP0023852 and Accelerate ACCESS  PHY220127.
Our neutron scattering experiments were performed at the Swiss Spallation Neutron Source SINQ, Paul Scherrer Institut, Switzerland, and the Spallation Neutron Source, a DOE Office of Science User Facility operated by the Oak Ridge National Laboratory, USA.
P.B. thanks SNSF projects 200021-188707 and 200020-182536 for financial support.
Certain commercial products or company names are identified here to describe our study adequately.
Such identification is not intended to imply recommendation or endorsement by the National Institute of Standards and Technology, nor is it intended to imply that the products or names identified are necessarily the best available for the purpose.




\bibliography{Fruhling_10jun2024}

\begin{thebibliography}{40}%
\makeatletter
\providecommand \@ifxundefined [1]{%
 \@ifx{#1\undefined}
}%
\providecommand \@ifnum [1]{%
 \ifnum #1\expandafter \@firstoftwo
 \else \expandafter \@secondoftwo
 \fi
}%
\providecommand \@ifx [1]{%
 \ifx #1\expandafter \@firstoftwo
 \else \expandafter \@secondoftwo
 \fi
}%
\providecommand \natexlab [1]{#1}%
\providecommand \enquote  [1]{``#1''}%
\providecommand \bibnamefont  [1]{#1}%
\providecommand \bibfnamefont [1]{#1}%
\providecommand \citenamefont [1]{#1}%
\providecommand \href@noop [0]{\@secondoftwo}%
\providecommand \href [0]{\begingroup \@sanitize@url \@href}%
\providecommand \@href[1]{\@@startlink{#1}\@@href}%
\providecommand \@@href[1]{\endgroup#1\@@endlink}%
\providecommand \@sanitize@url [0]{\catcode `\\12\catcode `\$12\catcode
  `\&12\catcode `\#12\catcode `\^12\catcode `\_12\catcode `\%12\relax}%
\providecommand \@@startlink[1]{}%
\providecommand \@@endlink[0]{}%
\providecommand \url  [0]{\begingroup\@sanitize@url \@url }%
\providecommand \@url [1]{\endgroup\@href {#1}{\urlprefix }}%
\providecommand \urlprefix  [0]{URL }%
\providecommand \Eprint [0]{\href }%
\providecommand \doibase [0]{https://doi.org/}%
\providecommand \selectlanguage [0]{\@gobble}%
\providecommand \bibinfo  [0]{\@secondoftwo}%
\providecommand \bibfield  [0]{\@secondoftwo}%
\providecommand \translation [1]{[#1]}%
\providecommand \BibitemOpen [0]{}%
\providecommand \bibitemStop [0]{}%
\providecommand \bibitemNoStop [0]{.\EOS\space}%
\providecommand \EOS [0]{\spacefactor3000\relax}%
\providecommand \BibitemShut  [1]{\csname bibitem#1\endcsname}%
\let\auto@bib@innerbib\@empty
\bibitem [{\citenamefont {Norman}(2016)}]{norman_colloquium_2016}%
  \BibitemOpen
  \bibfield  {author} {\bibinfo {author} {\bibfnamefont {M.}~\bibnamefont
  {Norman}},\ }\bibfield  {title} {\bibinfo {title} {Colloquium:
  {Herbertsmithite} and the search for the quantum spin liquid},\ }\href
  {https://doi.org/10.1103/RevModPhys.88.041002} {\bibfield  {journal}
  {\bibinfo  {journal} {Reviews of Modern Physics}\ }\textbf {\bibinfo {volume}
  {88}},\ \bibinfo {pages} {041002} (\bibinfo {year} {2016})}\BibitemShut
  {NoStop}%
\bibitem [{\citenamefont {Ghimire}\ and\ \citenamefont
  {Mazin}(2020)}]{Ghimire2020}%
  \BibitemOpen
  \bibfield  {author} {\bibinfo {author} {\bibfnamefont {N.~J.}\ \bibnamefont
  {Ghimire}}\ and\ \bibinfo {author} {\bibfnamefont {I.~I.}\ \bibnamefont
  {Mazin}},\ }\bibfield  {title} {\bibinfo {title} {Topology and correlations
  on the kagome lattice},\ }\href {https://doi.org/10.1038/s41563-019-0589-8}
  {\bibfield  {journal} {\bibinfo  {journal} {Nature Materials}\ }\textbf
  {\bibinfo {volume} {19}},\ \bibinfo {pages} {137} (\bibinfo {year}
  {2020})}\BibitemShut {NoStop}%
\bibitem [{\citenamefont {Tang}\ \emph {et~al.}(2011)\citenamefont {Tang},
  \citenamefont {Mei},\ and\ \citenamefont {Wen}}]{tang_high-temperature_2011}%
  \BibitemOpen
  \bibfield  {author} {\bibinfo {author} {\bibfnamefont {E.}~\bibnamefont
  {Tang}}, \bibinfo {author} {\bibfnamefont {J.-W.}\ \bibnamefont {Mei}},\ and\
  \bibinfo {author} {\bibfnamefont {X.-G.}\ \bibnamefont {Wen}},\ }\bibfield
  {title} {\bibinfo {title} {High-{Temperature} {Fractional} {Quantum} {Hall}
  {States}},\ }\href {https://doi.org/10.1103/PhysRevLett.106.236802}
  {\bibfield  {journal} {\bibinfo  {journal} {Physical Review Letters}\
  }\textbf {\bibinfo {volume} {106}},\ \bibinfo {pages} {236802} (\bibinfo
  {year} {2011})}\BibitemShut {NoStop}%
\bibitem [{\citenamefont {Lee}\ \emph {et~al.}(2023)\citenamefont {Lee},
  \citenamefont {Skomski}, \citenamefont {Wang}, \citenamefont {Orth},
  \citenamefont {Ren}, \citenamefont {Kang}, \citenamefont {Pathak},
  \citenamefont {Kutepov}, \citenamefont {Harmon}, \citenamefont {McQueeney},
  \citenamefont {Mazin},\ and\ \citenamefont {Ke}}]{lee_interplay_2023}%
  \BibitemOpen
  \bibfield  {author} {\bibinfo {author} {\bibfnamefont {Y.}~\bibnamefont
  {Lee}}, \bibinfo {author} {\bibfnamefont {R.}~\bibnamefont {Skomski}},
  \bibinfo {author} {\bibfnamefont {X.}~\bibnamefont {Wang}}, \bibinfo {author}
  {\bibfnamefont {P.~P.}\ \bibnamefont {Orth}}, \bibinfo {author}
  {\bibfnamefont {Y.}~\bibnamefont {Ren}}, \bibinfo {author} {\bibfnamefont
  {B.}~\bibnamefont {Kang}}, \bibinfo {author} {\bibfnamefont {A.~K.}\
  \bibnamefont {Pathak}}, \bibinfo {author} {\bibfnamefont {A.}~\bibnamefont
  {Kutepov}}, \bibinfo {author} {\bibfnamefont {B.~N.}\ \bibnamefont {Harmon}},
  \bibinfo {author} {\bibfnamefont {R.~J.}\ \bibnamefont {McQueeney}}, \bibinfo
  {author} {\bibfnamefont {I.~I.}\ \bibnamefont {Mazin}},\ and\ \bibinfo
  {author} {\bibfnamefont {L.}~\bibnamefont {Ke}},\ }\bibfield  {title}
  {\bibinfo {title} {Interplay between magnetism and band topology in {Kagome}
  magnets \ch{RMn6Sn6}},\ }\href {https://doi.org/10.1103/PhysRevB.108.045132}
  {\bibfield  {journal} {\bibinfo  {journal} {Physical Review B}\ }\textbf
  {\bibinfo {volume} {108}},\ \bibinfo {pages} {045132} (\bibinfo {year}
  {2023})}\BibitemShut {NoStop}%
\bibitem [{\citenamefont {Mazin}\ \emph {et~al.}(2014)\citenamefont {Mazin},
  \citenamefont {Jeschke}, \citenamefont {Lechermann}, \citenamefont {Lee},
  \citenamefont {Fink}, \citenamefont {Thomale},\ and\ \citenamefont
  {Valentí}}]{mazin_theoretical_2014}%
  \BibitemOpen
  \bibfield  {author} {\bibinfo {author} {\bibfnamefont {I.~I.}\ \bibnamefont
  {Mazin}}, \bibinfo {author} {\bibfnamefont {H.~O.}\ \bibnamefont {Jeschke}},
  \bibinfo {author} {\bibfnamefont {F.}~\bibnamefont {Lechermann}}, \bibinfo
  {author} {\bibfnamefont {H.}~\bibnamefont {Lee}}, \bibinfo {author}
  {\bibfnamefont {M.}~\bibnamefont {Fink}}, \bibinfo {author} {\bibfnamefont
  {R.}~\bibnamefont {Thomale}},\ and\ \bibinfo {author} {\bibfnamefont
  {R.}~\bibnamefont {Valentí}},\ }\bibfield  {title} {\bibinfo {title}
  {Theoretical prediction of a strongly correlated {Dirac} metal},\ }\href
  {https://doi.org/10.1038/ncomms5261} {\bibfield  {journal} {\bibinfo
  {journal} {Nature Communications}\ }\textbf {\bibinfo {volume} {5}},\
  \bibinfo {pages} {4261} (\bibinfo {year} {2014})}\BibitemShut {NoStop}%
\bibitem [{\citenamefont {Bolens}\ and\ \citenamefont
  {Nagaosa}(2019)}]{bolens_topological_2019}%
  \BibitemOpen
  \bibfield  {author} {\bibinfo {author} {\bibfnamefont {A.}~\bibnamefont
  {Bolens}}\ and\ \bibinfo {author} {\bibfnamefont {N.}~\bibnamefont
  {Nagaosa}},\ }\bibfield  {title} {\bibinfo {title} {Topological states on the
  breathing kagome lattice},\ }\href
  {https://doi.org/10.1103/PhysRevB.99.165141} {\bibfield  {journal} {\bibinfo
  {journal} {Physical Review B}\ }\textbf {\bibinfo {volume} {99}},\ \bibinfo
  {pages} {165141} (\bibinfo {year} {2019})}\BibitemShut {NoStop}%
\bibitem [{\citenamefont {Kang}\ \emph {et~al.}(2020)\citenamefont {Kang},
  \citenamefont {Ye}, \citenamefont {Fang}, \citenamefont {You}, \citenamefont
  {Levitan}, \citenamefont {Han}, \citenamefont {Facio}, \citenamefont
  {Jozwiak}, \citenamefont {Bostwick}, \citenamefont {Rotenberg}, \citenamefont
  {Chan}, \citenamefont {McDonald}, \citenamefont {Graf}, \citenamefont
  {Kaznatcheev}, \citenamefont {Vescovo}, \citenamefont {Bell}, \citenamefont
  {Kaxiras}, \citenamefont {van~den Brink}, \citenamefont {Richter},
  \citenamefont {Prasad~Ghimire}, \citenamefont {Checkelsky},\ and\
  \citenamefont {Comin}}]{kang_dirac_2020}%
  \BibitemOpen
  \bibfield  {author} {\bibinfo {author} {\bibfnamefont {M.}~\bibnamefont
  {Kang}}, \bibinfo {author} {\bibfnamefont {L.}~\bibnamefont {Ye}}, \bibinfo
  {author} {\bibfnamefont {S.}~\bibnamefont {Fang}}, \bibinfo {author}
  {\bibfnamefont {J.-S.}\ \bibnamefont {You}}, \bibinfo {author} {\bibfnamefont
  {A.}~\bibnamefont {Levitan}}, \bibinfo {author} {\bibfnamefont
  {M.}~\bibnamefont {Han}}, \bibinfo {author} {\bibfnamefont {J.~I.}\
  \bibnamefont {Facio}}, \bibinfo {author} {\bibfnamefont {C.}~\bibnamefont
  {Jozwiak}}, \bibinfo {author} {\bibfnamefont {A.}~\bibnamefont {Bostwick}},
  \bibinfo {author} {\bibfnamefont {E.}~\bibnamefont {Rotenberg}}, \bibinfo
  {author} {\bibfnamefont {M.~K.}\ \bibnamefont {Chan}}, \bibinfo {author}
  {\bibfnamefont {R.~D.}\ \bibnamefont {McDonald}}, \bibinfo {author}
  {\bibfnamefont {D.}~\bibnamefont {Graf}}, \bibinfo {author} {\bibfnamefont
  {K.}~\bibnamefont {Kaznatcheev}}, \bibinfo {author} {\bibfnamefont
  {E.}~\bibnamefont {Vescovo}}, \bibinfo {author} {\bibfnamefont {D.~C.}\
  \bibnamefont {Bell}}, \bibinfo {author} {\bibfnamefont {E.}~\bibnamefont
  {Kaxiras}}, \bibinfo {author} {\bibfnamefont {J.}~\bibnamefont {van~den
  Brink}}, \bibinfo {author} {\bibfnamefont {M.}~\bibnamefont {Richter}},
  \bibinfo {author} {\bibfnamefont {M.}~\bibnamefont {Prasad~Ghimire}},
  \bibinfo {author} {\bibfnamefont {J.~G.}\ \bibnamefont {Checkelsky}},\ and\
  \bibinfo {author} {\bibfnamefont {R.}~\bibnamefont {Comin}},\ }\bibfield
  {title} {\bibinfo {title} {Dirac fermions and flat bands in the ideal kagome
  metal {FeSn}},\ }\href {https://doi.org/10.1038/s41563-019-0531-0} {\bibfield
   {journal} {\bibinfo  {journal} {Nature Materials}\ }\textbf {\bibinfo
  {volume} {19}},\ \bibinfo {pages} {163} (\bibinfo {year} {2020})}\BibitemShut
  {NoStop}%
\bibitem [{\citenamefont {Venturini}\ \emph {et~al.}(1991)\citenamefont
  {Venturini}, \citenamefont {Idrissi},\ and\ \citenamefont
  {Malaman}}]{venturini_magnetic_1991}%
  \BibitemOpen
  \bibfield  {author} {\bibinfo {author} {\bibfnamefont {G.}~\bibnamefont
  {Venturini}}, \bibinfo {author} {\bibfnamefont {B.~C.~E.}\ \bibnamefont
  {Idrissi}},\ and\ \bibinfo {author} {\bibfnamefont {B.}~\bibnamefont
  {Malaman}},\ }\bibfield  {title} {\bibinfo {title} {Magnetic properties of
  \ch{RMn6Sn6} ({R} = {Sc}, {Y}, {Gd}-{Tm}, {Lu}) compounds with \ch{HfFe6Ge6}
  type structure},\ }\href {https://doi.org/10.1016/0304-8853(91)90108-M}
  {\bibfield  {journal} {\bibinfo  {journal} {Journal of Magnetism and Magnetic
  Materials}\ }\textbf {\bibinfo {volume} {94}},\ \bibinfo {pages} {35}
  (\bibinfo {year} {1991})}\BibitemShut {NoStop}%
\bibitem [{\citenamefont {El~Idrissi}\ \emph {et~al.}(1991)\citenamefont
  {El~Idrissi}, \citenamefont {Venturini}, \citenamefont {Malaman},\ and\
  \citenamefont {Fruchart}}]{el_idrissi_magnetic_1991}%
  \BibitemOpen
  \bibfield  {author} {\bibinfo {author} {\bibfnamefont {B.~C.}\ \bibnamefont
  {El~Idrissi}}, \bibinfo {author} {\bibfnamefont {G.}~\bibnamefont
  {Venturini}}, \bibinfo {author} {\bibfnamefont {B.}~\bibnamefont {Malaman}},\
  and\ \bibinfo {author} {\bibfnamefont {D.}~\bibnamefont {Fruchart}},\
  }\bibfield  {title} {\bibinfo {title} {Magnetic structures of
  {TbMn}$_6${Sn}$_6$ and {HoMn}$_6${Sn}$_6$ compounds from neutron diffraction
  study},\ }\href {https://doi.org/10.1016/0022-5088(91)90359-C} {\bibfield
  {journal} {\bibinfo  {journal} {Journal of the Less Common Metals}\ }\textbf
  {\bibinfo {volume} {175}},\ \bibinfo {pages} {143} (\bibinfo {year}
  {1991})}\BibitemShut {NoStop}%
\bibitem [{\citenamefont {Gorbunov}\ \emph {et~al.}(2012)\citenamefont
  {Gorbunov}, \citenamefont {Kuz’min}, \citenamefont {Uhlířová},
  \citenamefont {Žáček}, \citenamefont {Richter}, \citenamefont {Skourski},\
  and\ \citenamefont {Andreev}}]{gorbunov_magnetic_2012}%
  \BibitemOpen
  \bibfield  {author} {\bibinfo {author} {\bibfnamefont {D.~I.}\ \bibnamefont
  {Gorbunov}}, \bibinfo {author} {\bibfnamefont {M.~D.}\ \bibnamefont
  {Kuz’min}}, \bibinfo {author} {\bibfnamefont {K.}~\bibnamefont
  {Uhlířová}}, \bibinfo {author} {\bibfnamefont {M.}~\bibnamefont
  {Žáček}}, \bibinfo {author} {\bibfnamefont {M.}~\bibnamefont {Richter}},
  \bibinfo {author} {\bibfnamefont {Y.}~\bibnamefont {Skourski}},\ and\
  \bibinfo {author} {\bibfnamefont {A.~V.}\ \bibnamefont {Andreev}},\
  }\bibfield  {title} {\bibinfo {title} {Magnetic properties of a \ch{GdMn6Sn6}
  single crystal},\ }\href {https://doi.org/10.1016/j.jallcom.2011.12.016}
  {\bibfield  {journal} {\bibinfo  {journal} {Journal of Alloys and Compounds}\
  }\textbf {\bibinfo {volume} {519}},\ \bibinfo {pages} {47} (\bibinfo {year}
  {2012})}\BibitemShut {NoStop}%
\bibitem [{\citenamefont {Xu}\ \emph {et~al.}(2022)\citenamefont {Xu},
  \citenamefont {Yin}, \citenamefont {Ma}, \citenamefont {Tien}, \citenamefont
  {Qiang}, \citenamefont {Reddy}, \citenamefont {Zhou}, \citenamefont {Shen},
  \citenamefont {Lu}, \citenamefont {Chang}, \citenamefont {Qu},\ and\
  \citenamefont {Jia}}]{xu_topological_2022}%
  \BibitemOpen
  \bibfield  {author} {\bibinfo {author} {\bibfnamefont {X.}~\bibnamefont
  {Xu}}, \bibinfo {author} {\bibfnamefont {J.-X.}\ \bibnamefont {Yin}},
  \bibinfo {author} {\bibfnamefont {W.}~\bibnamefont {Ma}}, \bibinfo {author}
  {\bibfnamefont {H.-J.}\ \bibnamefont {Tien}}, \bibinfo {author}
  {\bibfnamefont {X.-B.}\ \bibnamefont {Qiang}}, \bibinfo {author}
  {\bibfnamefont {P.~V.~S.}\ \bibnamefont {Reddy}}, \bibinfo {author}
  {\bibfnamefont {H.}~\bibnamefont {Zhou}}, \bibinfo {author} {\bibfnamefont
  {J.}~\bibnamefont {Shen}}, \bibinfo {author} {\bibfnamefont {H.-Z.}\
  \bibnamefont {Lu}}, \bibinfo {author} {\bibfnamefont {T.-R.}\ \bibnamefont
  {Chang}}, \bibinfo {author} {\bibfnamefont {Z.}~\bibnamefont {Qu}},\ and\
  \bibinfo {author} {\bibfnamefont {S.}~\bibnamefont {Jia}},\ }\bibfield
  {title} {\bibinfo {title} {Topological charge-entropy scaling in kagome
  {Chern} magnet \ch{TbMn6Sn6}},\ }\href
  {https://doi.org/10.1038/s41467-022-28796-6} {\bibfield  {journal} {\bibinfo
  {journal} {Nature Communications}\ }\textbf {\bibinfo {volume} {13}},\
  \bibinfo {pages} {1197} (\bibinfo {year} {2022})}\BibitemShut {NoStop}%
\bibitem [{\citenamefont {Riberolles}\ \emph {et~al.}(2022)\citenamefont
  {Riberolles}, \citenamefont {Slade}, \citenamefont {Abernathy}, \citenamefont
  {Granroth}, \citenamefont {Li}, \citenamefont {Lee}, \citenamefont
  {Canfield}, \citenamefont {Ueland}, \citenamefont {Ke},\ and\ \citenamefont
  {McQueeney}}]{riberolles_low-temperature_2022}%
  \BibitemOpen
  \bibfield  {author} {\bibinfo {author} {\bibfnamefont {S.}~\bibnamefont
  {Riberolles}}, \bibinfo {author} {\bibfnamefont {T.~J.}\ \bibnamefont
  {Slade}}, \bibinfo {author} {\bibfnamefont {D.}~\bibnamefont {Abernathy}},
  \bibinfo {author} {\bibfnamefont {G.}~\bibnamefont {Granroth}}, \bibinfo
  {author} {\bibfnamefont {B.}~\bibnamefont {Li}}, \bibinfo {author}
  {\bibfnamefont {Y.}~\bibnamefont {Lee}}, \bibinfo {author} {\bibfnamefont
  {P.}~\bibnamefont {Canfield}}, \bibinfo {author} {\bibfnamefont
  {B.}~\bibnamefont {Ueland}}, \bibinfo {author} {\bibfnamefont
  {L.}~\bibnamefont {Ke}},\ and\ \bibinfo {author} {\bibfnamefont
  {R.}~\bibnamefont {McQueeney}},\ }\bibfield  {title} {\bibinfo {title}
  {Low-{Temperature} {Competing} {Magnetic} {Energy} {Scales} in the
  {Topological} {Ferrimagnet} \ch{TbMn6Sn6}},\ }\href
  {https://doi.org/10.1103/PhysRevX.12.021043} {\bibfield  {journal} {\bibinfo
  {journal} {Physical Review X}\ }\textbf {\bibinfo {volume} {12}},\ \bibinfo
  {pages} {021043} (\bibinfo {year} {2022})}\BibitemShut {NoStop}%
\bibitem [{\citenamefont {Mielke~III}\ \emph {et~al.}(2022)\citenamefont
  {Mielke~III}, \citenamefont {Ma}, \citenamefont {Pomjakushin}, \citenamefont
  {Zaharko}, \citenamefont {Sturniolo}, \citenamefont {Liu}, \citenamefont
  {Ukleev}, \citenamefont {White}, \citenamefont {Yin}, \citenamefont
  {Tsirkin}, \citenamefont {Larsen}, \citenamefont {Cochran}, \citenamefont
  {Medarde}, \citenamefont {Porée}, \citenamefont {Das}, \citenamefont
  {Gupta}, \citenamefont {Wang}, \citenamefont {Chang}, \citenamefont {Wang},
  \citenamefont {Khasanov}, \citenamefont {Neupert}, \citenamefont {Amato},
  \citenamefont {Liborio}, \citenamefont {Jia}, \citenamefont {Hasan},
  \citenamefont {Luetkens},\ and\ \citenamefont
  {Guguchia}}]{mielke_iii_low-temperature_2022}%
  \BibitemOpen
  \bibfield  {author} {\bibinfo {author} {\bibfnamefont {C.}~\bibnamefont
  {Mielke~III}}, \bibinfo {author} {\bibfnamefont {W.~L.}\ \bibnamefont {Ma}},
  \bibinfo {author} {\bibfnamefont {V.}~\bibnamefont {Pomjakushin}}, \bibinfo
  {author} {\bibfnamefont {O.}~\bibnamefont {Zaharko}}, \bibinfo {author}
  {\bibfnamefont {S.}~\bibnamefont {Sturniolo}}, \bibinfo {author}
  {\bibfnamefont {X.}~\bibnamefont {Liu}}, \bibinfo {author} {\bibfnamefont
  {V.}~\bibnamefont {Ukleev}}, \bibinfo {author} {\bibfnamefont {J.~S.}\
  \bibnamefont {White}}, \bibinfo {author} {\bibfnamefont {J.-X.}\ \bibnamefont
  {Yin}}, \bibinfo {author} {\bibfnamefont {S.~S.}\ \bibnamefont {Tsirkin}},
  \bibinfo {author} {\bibfnamefont {C.~B.}\ \bibnamefont {Larsen}}, \bibinfo
  {author} {\bibfnamefont {T.~A.}\ \bibnamefont {Cochran}}, \bibinfo {author}
  {\bibfnamefont {M.}~\bibnamefont {Medarde}}, \bibinfo {author} {\bibfnamefont
  {V.}~\bibnamefont {Porée}}, \bibinfo {author} {\bibfnamefont
  {D.}~\bibnamefont {Das}}, \bibinfo {author} {\bibfnamefont {R.}~\bibnamefont
  {Gupta}}, \bibinfo {author} {\bibfnamefont {C.~N.}\ \bibnamefont {Wang}},
  \bibinfo {author} {\bibfnamefont {J.}~\bibnamefont {Chang}}, \bibinfo
  {author} {\bibfnamefont {Z.~Q.}\ \bibnamefont {Wang}}, \bibinfo {author}
  {\bibfnamefont {R.}~\bibnamefont {Khasanov}}, \bibinfo {author}
  {\bibfnamefont {T.}~\bibnamefont {Neupert}}, \bibinfo {author} {\bibfnamefont
  {A.}~\bibnamefont {Amato}}, \bibinfo {author} {\bibfnamefont
  {L.}~\bibnamefont {Liborio}}, \bibinfo {author} {\bibfnamefont
  {S.}~\bibnamefont {Jia}}, \bibinfo {author} {\bibfnamefont {M.~Z.}\
  \bibnamefont {Hasan}}, \bibinfo {author} {\bibfnamefont {H.}~\bibnamefont
  {Luetkens}},\ and\ \bibinfo {author} {\bibfnamefont {Z.}~\bibnamefont
  {Guguchia}},\ }\bibfield  {title} {\bibinfo {title} {Low-temperature magnetic
  crossover in the topological kagome magnet {TbMn}$_6${Sn}$_6$},\ }\href
  {https://doi.org/10.1038/s42005-022-00885-4} {\bibfield  {journal} {\bibinfo
  {journal} {Communications Physics}\ }\textbf {\bibinfo {volume} {5}},\
  \bibinfo {pages} {1} (\bibinfo {year} {2022})}\BibitemShut {NoStop}%
\bibitem [{\citenamefont {Jones}\ \emph {et~al.}(2022)\citenamefont {Jones},
  \citenamefont {Das}, \citenamefont {Bhandari}, \citenamefont {Liu},
  \citenamefont {Siegfried}, \citenamefont {Ghimire}, \citenamefont {Tsirkin},
  \citenamefont {Mazin},\ and\ \citenamefont {Ghimire}}]{jones_origin_2022}%
  \BibitemOpen
  \bibfield  {author} {\bibinfo {author} {\bibfnamefont {D.~C.}\ \bibnamefont
  {Jones}}, \bibinfo {author} {\bibfnamefont {S.}~\bibnamefont {Das}}, \bibinfo
  {author} {\bibfnamefont {H.}~\bibnamefont {Bhandari}}, \bibinfo {author}
  {\bibfnamefont {X.}~\bibnamefont {Liu}}, \bibinfo {author} {\bibfnamefont
  {P.}~\bibnamefont {Siegfried}}, \bibinfo {author} {\bibfnamefont {M.~P.}\
  \bibnamefont {Ghimire}}, \bibinfo {author} {\bibfnamefont {S.~S.}\
  \bibnamefont {Tsirkin}}, \bibinfo {author} {\bibfnamefont {I.~I.}\
  \bibnamefont {Mazin}},\ and\ \bibinfo {author} {\bibfnamefont {N.~J.}\
  \bibnamefont {Ghimire}},\ }\href {https://doi.org/10.48550/arXiv.2203.17246}
  {\bibinfo {title} {Origin of spin reorientation and intrinsic anomalous
  {Hall} effect in the kagome ferrimagnet \ch{TbMn6Sn6}}} (\bibinfo {year}
  {2022}),\ \bibinfo {note} {arXiv:2203.17246 [cond-mat]}\BibitemShut {NoStop}%
\bibitem [{\citenamefont {Dally}\ \emph {et~al.}(2021)\citenamefont {Dally},
  \citenamefont {Lynn}, \citenamefont {Ghimire}, \citenamefont {Michel},
  \citenamefont {Siegfried},\ and\ \citenamefont {Mazin}}]{dally_chiral_2021}%
  \BibitemOpen
  \bibfield  {author} {\bibinfo {author} {\bibfnamefont {R.~L.}\ \bibnamefont
  {Dally}}, \bibinfo {author} {\bibfnamefont {J.~W.}\ \bibnamefont {Lynn}},
  \bibinfo {author} {\bibfnamefont {N.~J.}\ \bibnamefont {Ghimire}}, \bibinfo
  {author} {\bibfnamefont {D.}~\bibnamefont {Michel}}, \bibinfo {author}
  {\bibfnamefont {P.}~\bibnamefont {Siegfried}},\ and\ \bibinfo {author}
  {\bibfnamefont {I.~I.}\ \bibnamefont {Mazin}},\ }\bibfield  {title} {\bibinfo
  {title} {Chiral properties of the zero-field spiral state and field-induced
  magnetic phases of the itinerant kagome metal \ch{YMn6Sn6}},\ }\href
  {https://doi.org/10.1103/PhysRevB.103.094413} {\bibfield  {journal} {\bibinfo
   {journal} {Physical Review B}\ }\textbf {\bibinfo {volume} {103}},\ \bibinfo
  {pages} {094413} (\bibinfo {year} {2021})}\BibitemShut {NoStop}%
\bibitem [{\citenamefont {Ghimire}\ \emph {et~al.}(2020)\citenamefont
  {Ghimire}, \citenamefont {Dally}, \citenamefont {Poudel}, \citenamefont
  {Jones}, \citenamefont {Michel}, \citenamefont {Magar}, \citenamefont
  {Bleuel}, \citenamefont {McGuire}, \citenamefont {Jiang}, \citenamefont
  {Mitchell}, \citenamefont {Lynn},\ and\ \citenamefont
  {Mazin}}]{ghimire_competing_2020}%
  \BibitemOpen
  \bibfield  {author} {\bibinfo {author} {\bibfnamefont {N.~J.}\ \bibnamefont
  {Ghimire}}, \bibinfo {author} {\bibfnamefont {R.~L.}\ \bibnamefont {Dally}},
  \bibinfo {author} {\bibfnamefont {L.}~\bibnamefont {Poudel}}, \bibinfo
  {author} {\bibfnamefont {D.~C.}\ \bibnamefont {Jones}}, \bibinfo {author}
  {\bibfnamefont {D.}~\bibnamefont {Michel}}, \bibinfo {author} {\bibfnamefont
  {N.~T.}\ \bibnamefont {Magar}}, \bibinfo {author} {\bibfnamefont
  {M.}~\bibnamefont {Bleuel}}, \bibinfo {author} {\bibfnamefont {M.~A.}\
  \bibnamefont {McGuire}}, \bibinfo {author} {\bibfnamefont {J.~S.}\
  \bibnamefont {Jiang}}, \bibinfo {author} {\bibfnamefont {J.~F.}\ \bibnamefont
  {Mitchell}}, \bibinfo {author} {\bibfnamefont {J.~W.}\ \bibnamefont {Lynn}},\
  and\ \bibinfo {author} {\bibfnamefont {I.~I.}\ \bibnamefont {Mazin}},\
  }\bibfield  {title} {\bibinfo {title} {Competing magnetic phases and
  fluctuation-driven scalar spin chirality in the kagome metal \ch{YMn6Sn6}},\
  }\href {https://doi.org/10.1126/sciadv.abe2680} {\bibfield  {journal}
  {\bibinfo  {journal} {Science Advances}\ }\textbf {\bibinfo {volume} {6}},\
  \bibinfo {pages} {eabe2680} (\bibinfo {year} {2020})}\BibitemShut {NoStop}%
\bibitem [{\citenamefont {Zhang}\ \emph {et~al.}(2022)\citenamefont {Zhang},
  \citenamefont {Liu}, \citenamefont {Zhang}, \citenamefont {Hou},
  \citenamefont {Fu}, \citenamefont {Zhang}, \citenamefont {Gao},\ and\
  \citenamefont {Liu}}]{zhang_magnetic_2022}%
  \BibitemOpen
  \bibfield  {author} {\bibinfo {author} {\bibfnamefont {H.}~\bibnamefont
  {Zhang}}, \bibinfo {author} {\bibfnamefont {C.}~\bibnamefont {Liu}}, \bibinfo
  {author} {\bibfnamefont {Y.}~\bibnamefont {Zhang}}, \bibinfo {author}
  {\bibfnamefont {Z.}~\bibnamefont {Hou}}, \bibinfo {author} {\bibfnamefont
  {X.}~\bibnamefont {Fu}}, \bibinfo {author} {\bibfnamefont {X.}~\bibnamefont
  {Zhang}}, \bibinfo {author} {\bibfnamefont {X.}~\bibnamefont {Gao}},\ and\
  \bibinfo {author} {\bibfnamefont {J.}~\bibnamefont {Liu}},\ }\bibfield
  {title} {\bibinfo {title} {Magnetic field-induced nontrivial spin chirality
  and large topological {Hall} effect in kagome magnet \ch{ScMn6Sn6}},\ }\href
  {https://doi.org/10.1063/5.0127091} {\bibfield  {journal} {\bibinfo
  {journal} {Applied Physics Letters}\ }\textbf {\bibinfo {volume} {121}},\
  \bibinfo {pages} {202401} (\bibinfo {year} {2022})}\BibitemShut {NoStop}%
\bibitem [{\citenamefont {Casey}\ \emph {et~al.}(2023)\citenamefont {Casey},
  \citenamefont {Samatham}, \citenamefont {Burgio}, \citenamefont {Kramer},
  \citenamefont {Sawon}, \citenamefont {Huff},\ and\ \citenamefont
  {Pathak}}]{casey_spin-flop_2023}%
  \BibitemOpen
  \bibfield  {author} {\bibinfo {author} {\bibfnamefont {J.}~\bibnamefont
  {Casey}}, \bibinfo {author} {\bibfnamefont {S.~S.}\ \bibnamefont {Samatham}},
  \bibinfo {author} {\bibfnamefont {C.}~\bibnamefont {Burgio}}, \bibinfo
  {author} {\bibfnamefont {N.}~\bibnamefont {Kramer}}, \bibinfo {author}
  {\bibfnamefont {A.}~\bibnamefont {Sawon}}, \bibinfo {author} {\bibfnamefont
  {J.}~\bibnamefont {Huff}},\ and\ \bibinfo {author} {\bibfnamefont {A.~K.}\
  \bibnamefont {Pathak}},\ }\bibfield  {title} {\bibinfo {title} {Spin-flop
  quasi metamagnetic, anisotropic magnetic, and electrical transport behavior
  of {Ho} substituted kagome magnet \ch{ErMn6Sn6}},\ }\href
  {https://doi.org/10.1103/PhysRevMaterials.7.074402} {\bibfield  {journal}
  {\bibinfo  {journal} {Physical Review Materials}\ }\textbf {\bibinfo {volume}
  {7}},\ \bibinfo {pages} {074402} (\bibinfo {year} {2023})}\BibitemShut
  {NoStop}%
\bibitem [{sup()}]{suppmatt}%
  \BibitemOpen
  \href {https://journals.aps.org} {}\bibinfo {note} {See the Supplemental
  Material for details of crystal growth, magnetic and transport
  characterizations, neutron scattering, and first-principles
  calculations.}\BibitemShut {Stop}%
\bibitem [{\citenamefont
  {Rodr\'{i}guez-Carvajal}(1993)}]{rodriguez-carvajal_recent_1993}%
  \BibitemOpen
  \bibfield  {author} {\bibinfo {author} {\bibfnamefont {J.}~\bibnamefont
  {Rodr\'{i}guez-Carvajal}},\ }\bibfield  {title} {\bibinfo {title} {Recent
  advances in magnetic structure determination by neutron powder diffraction},\
  }\href {https://doi.org/10.1016/0921-4526(93)90108-I} {\bibfield  {journal}
  {\bibinfo  {journal} {Physica B: Condensed Matter}\ }\textbf {\bibinfo
  {volume} {192}},\ \bibinfo {pages} {55} (\bibinfo {year} {1993})}\BibitemShut
  {NoStop}%
\bibitem [{\citenamefont {Momma}\ and\ \citenamefont
  {Izumi}(2011)}]{momma_vesta_2011}%
  \BibitemOpen
  \bibfield  {author} {\bibinfo {author} {\bibfnamefont {K.}~\bibnamefont
  {Momma}}\ and\ \bibinfo {author} {\bibfnamefont {F.}~\bibnamefont {Izumi}},\
  }\bibfield  {title} {\bibinfo {title} {{VESTA} 3 for three-dimensional
  visualization of crystal, volumetric and morphology data},\ }\href
  {https://doi.org/10.1107/S0021889811038970} {\bibfield  {journal} {\bibinfo
  {journal} {Journal of Applied Crystallography}\ }\textbf {\bibinfo {volume}
  {44}},\ \bibinfo {pages} {1272} (\bibinfo {year} {2011})}\BibitemShut
  {NoStop}%
\bibitem [{\citenamefont {Hohenberg}\ and\ \citenamefont
  {Kohn}(1964)}]{hohenberg1964inhomogeneous}%
  \BibitemOpen
  \bibfield  {author} {\bibinfo {author} {\bibfnamefont {P.}~\bibnamefont
  {Hohenberg}}\ and\ \bibinfo {author} {\bibfnamefont {W.}~\bibnamefont
  {Kohn}},\ }\bibfield  {title} {\bibinfo {title} {Inhomogeneous electron
  gas},\ }\href@noop {} {\bibfield  {journal} {\bibinfo  {journal} {Physical
  Review}\ }\textbf {\bibinfo {volume} {136}},\ \bibinfo {pages} {B864}
  (\bibinfo {year} {1964})}\BibitemShut {NoStop}%
\bibitem [{\citenamefont {Kresse}\ and\ \citenamefont
  {Furthm{\"u}ller}(1996)}]{kresse1996efficient}%
  \BibitemOpen
  \bibfield  {author} {\bibinfo {author} {\bibfnamefont {G.}~\bibnamefont
  {Kresse}}\ and\ \bibinfo {author} {\bibfnamefont {J.}~\bibnamefont
  {Furthm{\"u}ller}},\ }\bibfield  {title} {\bibinfo {title} {Efficient
  iterative schemes for ab initio total-energy calculations using a plane-wave
  basis set},\ }\href@noop {} {\bibfield  {journal} {\bibinfo  {journal}
  {Physical Review B}\ }\textbf {\bibinfo {volume} {54}},\ \bibinfo {pages}
  {11169} (\bibinfo {year} {1996})}\BibitemShut {NoStop}%
\bibitem [{\citenamefont {Kresse}\ and\ \citenamefont
  {Joubert}(1999)}]{kresse1999ultrasoft}%
  \BibitemOpen
  \bibfield  {author} {\bibinfo {author} {\bibfnamefont {G.}~\bibnamefont
  {Kresse}}\ and\ \bibinfo {author} {\bibfnamefont {D.}~\bibnamefont
  {Joubert}},\ }\bibfield  {title} {\bibinfo {title} {From ultrasoft
  pseudopotentials to the projector augmented-wave method},\ }\href@noop {}
  {\bibfield  {journal} {\bibinfo  {journal} {Physical Review B}\ }\textbf
  {\bibinfo {volume} {59}},\ \bibinfo {pages} {1758} (\bibinfo {year}
  {1999})}\BibitemShut {NoStop}%
\bibitem [{\citenamefont {Perdew}\ \emph {et~al.}(1996)\citenamefont {Perdew},
  \citenamefont {Burke},\ and\ \citenamefont
  {Ernzerhof}}]{perdew1996generalized}%
  \BibitemOpen
  \bibfield  {author} {\bibinfo {author} {\bibfnamefont {J.~P.}\ \bibnamefont
  {Perdew}}, \bibinfo {author} {\bibfnamefont {K.}~\bibnamefont {Burke}},\ and\
  \bibinfo {author} {\bibfnamefont {M.}~\bibnamefont {Ernzerhof}},\ }\bibfield
  {title} {\bibinfo {title} {Generalized gradient approximation made simple},\
  }\href@noop {} {\bibfield  {journal} {\bibinfo  {journal} {Physical Review
  Letters}\ }\textbf {\bibinfo {volume} {77}},\ \bibinfo {pages} {3865}
  (\bibinfo {year} {1996})}\BibitemShut {NoStop}%
\bibitem [{\citenamefont {Monkhorst}\ and\ \citenamefont
  {Pack}(1976)}]{Monkhorst1976}%
  \BibitemOpen
  \bibfield  {author} {\bibinfo {author} {\bibfnamefont {H.~J.}\ \bibnamefont
  {Monkhorst}}\ and\ \bibinfo {author} {\bibfnamefont {J.~D.}\ \bibnamefont
  {Pack}},\ }\bibfield  {title} {\bibinfo {title} {Special points for
  brillouin-zone integrations},\ }\href
  {https://doi.org/10.1103/PhysRevB.13.5188} {\bibfield  {journal} {\bibinfo
  {journal} {Phys. Rev. B}\ }\textbf {\bibinfo {volume} {13}},\ \bibinfo
  {pages} {5188} (\bibinfo {year} {1976})}\BibitemShut {NoStop}%
\bibitem [{\citenamefont {Allen}\ and\ \citenamefont
  {Watson}(2014)}]{allen2014occu}%
  \BibitemOpen
  \bibfield  {author} {\bibinfo {author} {\bibfnamefont {J.~P.}\ \bibnamefont
  {Allen}}\ and\ \bibinfo {author} {\bibfnamefont {G.~W.}\ \bibnamefont
  {Watson}},\ }\bibfield  {title} {\bibinfo {title} {Occupation matrix control
  of d- and f-electron localisations using dft + u},\ }\href
  {https://doi.org/10.1039/C4CP01083C} {\bibfield  {journal} {\bibinfo
  {journal} {Phys. Chem. Chem. Phys.}\ }\textbf {\bibinfo {volume} {16}},\
  \bibinfo {pages} {21016} (\bibinfo {year} {2014})}\BibitemShut {NoStop}%
\bibitem [{\citenamefont {Coates}\ \emph {et~al.}(2018)\citenamefont {Coates},
  \citenamefont {Cao}, \citenamefont {Chakoumakos}, \citenamefont {Frontzek},
  \citenamefont {Hoffmann}, \citenamefont {Kovalevsky}, \citenamefont {Liu},
  \citenamefont {Meilleur}, \citenamefont {dos Santos}, \citenamefont {Myles},
  \citenamefont {Wang},\ and\ \citenamefont {Ye}}]{coates_suite-level_2018}%
  \BibitemOpen
  \bibfield  {author} {\bibinfo {author} {\bibfnamefont {L.}~\bibnamefont
  {Coates}}, \bibinfo {author} {\bibfnamefont {H.~B.}\ \bibnamefont {Cao}},
  \bibinfo {author} {\bibfnamefont {B.~C.}\ \bibnamefont {Chakoumakos}},
  \bibinfo {author} {\bibfnamefont {M.~D.}\ \bibnamefont {Frontzek}}, \bibinfo
  {author} {\bibfnamefont {C.}~\bibnamefont {Hoffmann}}, \bibinfo {author}
  {\bibfnamefont {A.~Y.}\ \bibnamefont {Kovalevsky}}, \bibinfo {author}
  {\bibfnamefont {Y.}~\bibnamefont {Liu}}, \bibinfo {author} {\bibfnamefont
  {F.}~\bibnamefont {Meilleur}}, \bibinfo {author} {\bibfnamefont {A.~M.}\
  \bibnamefont {dos Santos}}, \bibinfo {author} {\bibfnamefont {D.~A.~A.}\
  \bibnamefont {Myles}}, \bibinfo {author} {\bibfnamefont {X.~P.}\ \bibnamefont
  {Wang}},\ and\ \bibinfo {author} {\bibfnamefont {F.}~\bibnamefont {Ye}},\
  }\bibfield  {title} {\bibinfo {title} {A suite-level review of the neutron
  single-crystal diffraction instruments at {Oak} {Ridge} {National}
  {Laboratory}},\ }\href {https://doi.org/10.1063/1.5030896} {\bibfield
  {journal} {\bibinfo  {journal} {Review of Scientific Instruments}\ }\textbf
  {\bibinfo {volume} {89}},\ \bibinfo {pages} {092802} (\bibinfo {year}
  {2018})}\BibitemShut {NoStop}%
\bibitem [{\citenamefont {Zikovsky}\ \emph {et~al.}(2011)\citenamefont
  {Zikovsky}, \citenamefont {Peterson}, \citenamefont {Wang}, \citenamefont
  {Frost},\ and\ \citenamefont {Hoffmann}}]{zikovsky_crystalplan_2011}%
  \BibitemOpen
  \bibfield  {author} {\bibinfo {author} {\bibfnamefont {J.}~\bibnamefont
  {Zikovsky}}, \bibinfo {author} {\bibfnamefont {P.~F.}\ \bibnamefont
  {Peterson}}, \bibinfo {author} {\bibfnamefont {X.~P.}\ \bibnamefont {Wang}},
  \bibinfo {author} {\bibfnamefont {M.}~\bibnamefont {Frost}},\ and\ \bibinfo
  {author} {\bibfnamefont {C.}~\bibnamefont {Hoffmann}},\ }\bibfield  {title}
  {\bibinfo {title} {{CrystalPlan}: an experiment-planning tool for
  crystallography},\ }\href {https://doi.org/10.1107/S0021889811007102}
  {\bibfield  {journal} {\bibinfo  {journal} {Journal of Applied
  Crystallography}\ }\textbf {\bibinfo {volume} {44}},\ \bibinfo {pages} {418}
  (\bibinfo {year} {2011})}\BibitemShut {NoStop}%
\bibitem [{\citenamefont {Schultz}\ \emph {et~al.}(2014)\citenamefont
  {Schultz}, \citenamefont {Jørgensen}, \citenamefont {Wang}, \citenamefont
  {Mikkelson}, \citenamefont {Mikkelson}, \citenamefont {Lynch}, \citenamefont
  {Peterson}, \citenamefont {Green},\ and\ \citenamefont
  {Hoffmann}}]{schultz_integration_2014}%
  \BibitemOpen
  \bibfield  {author} {\bibinfo {author} {\bibfnamefont {A.~J.}\ \bibnamefont
  {Schultz}}, \bibinfo {author} {\bibfnamefont {M.~R.~V.}\ \bibnamefont
  {Jørgensen}}, \bibinfo {author} {\bibfnamefont {X.}~\bibnamefont {Wang}},
  \bibinfo {author} {\bibfnamefont {R.~L.}\ \bibnamefont {Mikkelson}}, \bibinfo
  {author} {\bibfnamefont {D.~J.}\ \bibnamefont {Mikkelson}}, \bibinfo {author}
  {\bibfnamefont {V.~E.}\ \bibnamefont {Lynch}}, \bibinfo {author}
  {\bibfnamefont {P.~F.}\ \bibnamefont {Peterson}}, \bibinfo {author}
  {\bibfnamefont {M.~L.}\ \bibnamefont {Green}},\ and\ \bibinfo {author}
  {\bibfnamefont {C.~M.}\ \bibnamefont {Hoffmann}},\ }\bibfield  {title}
  {\bibinfo {title} {Integration of neutron time-of-flight single-crystal
  {Bragg} peaks in reciprocal space},\ }\href
  {https://doi.org/10.1107/S1600576714006372} {\bibfield  {journal} {\bibinfo
  {journal} {Journal of Applied Crystallography}\ }\textbf {\bibinfo {volume}
  {47}},\ \bibinfo {pages} {915} (\bibinfo {year} {2014})},\ \bibinfo {note}
  {publisher: International Union of Crystallography}\BibitemShut {NoStop}%
\bibitem [{\citenamefont {Petříček}\ \emph {et~al.}(2023)\citenamefont
  {Petříček}, \citenamefont {Palatinus}, \citenamefont {Plášil},\ and\
  \citenamefont {Dušek}}]{Petricek_Jana_2023}%
  \BibitemOpen
  \bibfield  {author} {\bibinfo {author} {\bibfnamefont {V.}~\bibnamefont
  {Petříček}}, \bibinfo {author} {\bibfnamefont {L.}~\bibnamefont
  {Palatinus}}, \bibinfo {author} {\bibfnamefont {J.}~\bibnamefont
  {Plášil}},\ and\ \bibinfo {author} {\bibfnamefont {M.}~\bibnamefont
  {Dušek}},\ }\bibfield  {title} {\bibinfo {title} {Jana2020 – a new
  version of the crystallographic computing system jana},\ }\href
  {https://doi.org/doi:10.1515/zkri-2023-0005} {\bibfield  {journal} {\bibinfo
  {journal} {Zeitschrift für Kristallographie - Crystalline Materials}\
  }\textbf {\bibinfo {volume} {238}},\ \bibinfo {pages} {271} (\bibinfo {year}
  {2023})}\BibitemShut {NoStop}%
\bibitem [{\citenamefont {Malaman}\ \emph {et~al.}(1999)\citenamefont
  {Malaman}, \citenamefont {Venturini}, \citenamefont {Welter}, \citenamefont
  {Sanchez}, \citenamefont {Vulliet},\ and\ \citenamefont
  {Ressouche}}]{malaman_magnetic_1999}%
  \BibitemOpen
  \bibfield  {author} {\bibinfo {author} {\bibfnamefont {B.}~\bibnamefont
  {Malaman}}, \bibinfo {author} {\bibfnamefont {G.}~\bibnamefont {Venturini}},
  \bibinfo {author} {\bibfnamefont {R.}~\bibnamefont {Welter}}, \bibinfo
  {author} {\bibfnamefont {J.~P.}\ \bibnamefont {Sanchez}}, \bibinfo {author}
  {\bibfnamefont {P.}~\bibnamefont {Vulliet}},\ and\ \bibinfo {author}
  {\bibfnamefont {E.}~\bibnamefont {Ressouche}},\ }\bibfield  {title} {\bibinfo
  {title} {Magnetic properties of \ch{RMn6Sn6} ({R}={Gd}–{Er}) compounds from
  neutron diffraction and {Mössbauer} measurements},\ }\href
  {https://doi.org/10.1016/S0304-8853(99)00300-5} {\bibfield  {journal}
  {\bibinfo  {journal} {Journal of Magnetism and Magnetic Materials}\ }\textbf
  {\bibinfo {volume} {202}},\ \bibinfo {pages} {519} (\bibinfo {year}
  {1999})}\BibitemShut {NoStop}%
\bibitem [{\citenamefont {Kurumaji}\ \emph {et~al.}(2019)\citenamefont
  {Kurumaji}, \citenamefont {Nakajima}, \citenamefont {Hirschberger},
  \citenamefont {Kikkawa}, \citenamefont {Yamasaki}, \citenamefont {Sagayama},
  \citenamefont {Nakao}, \citenamefont {Taguchi}, \citenamefont {Arima},\ and\
  \citenamefont {Tokura}}]{kurumaji_skyrmion_2019}%
  \BibitemOpen
  \bibfield  {author} {\bibinfo {author} {\bibfnamefont {T.}~\bibnamefont
  {Kurumaji}}, \bibinfo {author} {\bibfnamefont {T.}~\bibnamefont {Nakajima}},
  \bibinfo {author} {\bibfnamefont {M.}~\bibnamefont {Hirschberger}}, \bibinfo
  {author} {\bibfnamefont {A.}~\bibnamefont {Kikkawa}}, \bibinfo {author}
  {\bibfnamefont {Y.}~\bibnamefont {Yamasaki}}, \bibinfo {author}
  {\bibfnamefont {H.}~\bibnamefont {Sagayama}}, \bibinfo {author}
  {\bibfnamefont {H.}~\bibnamefont {Nakao}}, \bibinfo {author} {\bibfnamefont
  {Y.}~\bibnamefont {Taguchi}}, \bibinfo {author} {\bibfnamefont {T.-h.}\
  \bibnamefont {Arima}},\ and\ \bibinfo {author} {\bibfnamefont
  {Y.}~\bibnamefont {Tokura}},\ }\bibfield  {title} {\bibinfo {title} {Skyrmion
  lattice with a giant topological {Hall} effect in a frustrated
  triangular-lattice magnet},\ }\href {https://doi.org/10.1126/science.aau0968}
  {\bibfield  {journal} {\bibinfo  {journal} {Science}\ }\textbf {\bibinfo
  {volume} {365}},\ \bibinfo {pages} {914} (\bibinfo {year}
  {2019})}\BibitemShut {NoStop}%
\bibitem [{\citenamefont {Shao}\ \emph {et~al.}(2019)\citenamefont {Shao},
  \citenamefont {Liu}, \citenamefont {Yu}, \citenamefont {Kim}, \citenamefont
  {Che}, \citenamefont {Tang}, \citenamefont {He}, \citenamefont {Tserkovnyak},
  \citenamefont {Shi},\ and\ \citenamefont {Wang}}]{shao_topological_2019}%
  \BibitemOpen
  \bibfield  {author} {\bibinfo {author} {\bibfnamefont {Q.}~\bibnamefont
  {Shao}}, \bibinfo {author} {\bibfnamefont {Y.}~\bibnamefont {Liu}}, \bibinfo
  {author} {\bibfnamefont {G.}~\bibnamefont {Yu}}, \bibinfo {author}
  {\bibfnamefont {S.~K.}\ \bibnamefont {Kim}}, \bibinfo {author} {\bibfnamefont
  {X.}~\bibnamefont {Che}}, \bibinfo {author} {\bibfnamefont {C.}~\bibnamefont
  {Tang}}, \bibinfo {author} {\bibfnamefont {Q.~L.}\ \bibnamefont {He}},
  \bibinfo {author} {\bibfnamefont {Y.}~\bibnamefont {Tserkovnyak}}, \bibinfo
  {author} {\bibfnamefont {J.}~\bibnamefont {Shi}},\ and\ \bibinfo {author}
  {\bibfnamefont {K.~L.}\ \bibnamefont {Wang}},\ }\bibfield  {title} {\bibinfo
  {title} {Topological {Hall} effect at above room temperature in
  heterostructures composed of a magnetic insulator and a heavy metal},\ }\href
  {https://doi.org/10.1038/s41928-019-0246-x} {\bibfield  {journal} {\bibinfo
  {journal} {Nature Electronics}\ }\textbf {\bibinfo {volume} {2}},\ \bibinfo
  {pages} {182} (\bibinfo {year} {2019})}\BibitemShut {NoStop}%
\bibitem [{\citenamefont {Verma}\ \emph {et~al.}(2022)\citenamefont {Verma},
  \citenamefont {Addison},\ and\ \citenamefont
  {Randeria}}]{verma_unified_2022}%
  \BibitemOpen
  \bibfield  {author} {\bibinfo {author} {\bibfnamefont {N.}~\bibnamefont
  {Verma}}, \bibinfo {author} {\bibfnamefont {Z.}~\bibnamefont {Addison}},\
  and\ \bibinfo {author} {\bibfnamefont {M.}~\bibnamefont {Randeria}},\
  }\bibfield  {title} {\bibinfo {title} {Unified theory of the anomalous and
  topological {Hall} effects with phase-space {Berry} curvatures},\ }\href
  {https://doi.org/10.1126/sciadv.abq2765} {\bibfield  {journal} {\bibinfo
  {journal} {Science Advances}\ }\textbf {\bibinfo {volume} {8}},\ \bibinfo
  {pages} {eabq2765} (\bibinfo {year} {2022})}\BibitemShut {NoStop}%
\bibitem [{\citenamefont {Wang}\ \emph {et~al.}(2021)\citenamefont {Wang},
  \citenamefont {Neubauer}, \citenamefont {Duan}, \citenamefont {Yin},
  \citenamefont {Fujitsu}, \citenamefont {Hosono}, \citenamefont {Ye},
  \citenamefont {Zhang}, \citenamefont {Chi}, \citenamefont {Krycka},
  \citenamefont {Lei},\ and\ \citenamefont {Dai}}]{wang_field-induced_2021}%
  \BibitemOpen
  \bibfield  {author} {\bibinfo {author} {\bibfnamefont {Q.}~\bibnamefont
  {Wang}}, \bibinfo {author} {\bibfnamefont {K.~J.}\ \bibnamefont {Neubauer}},
  \bibinfo {author} {\bibfnamefont {C.}~\bibnamefont {Duan}}, \bibinfo {author}
  {\bibfnamefont {Q.}~\bibnamefont {Yin}}, \bibinfo {author} {\bibfnamefont
  {S.}~\bibnamefont {Fujitsu}}, \bibinfo {author} {\bibfnamefont
  {H.}~\bibnamefont {Hosono}}, \bibinfo {author} {\bibfnamefont
  {F.}~\bibnamefont {Ye}}, \bibinfo {author} {\bibfnamefont {R.}~\bibnamefont
  {Zhang}}, \bibinfo {author} {\bibfnamefont {S.}~\bibnamefont {Chi}}, \bibinfo
  {author} {\bibfnamefont {K.}~\bibnamefont {Krycka}}, \bibinfo {author}
  {\bibfnamefont {H.}~\bibnamefont {Lei}},\ and\ \bibinfo {author}
  {\bibfnamefont {P.}~\bibnamefont {Dai}},\ }\bibfield  {title} {\bibinfo
  {title} {Field-induced topological {Hall} effect and double-fan spin
  structure with a $c$-axis component in the metallic kagome antiferromagnetic
  compound \ch{YMn6Sn6}.},\ }\href
  {https://doi.org/10.1103/PhysRevB.103.014416} {\bibfield  {journal} {\bibinfo
   {journal} {Physical Review B}\ }\textbf {\bibinfo {volume} {103}},\ \bibinfo
  {pages} {014416} (\bibinfo {year} {2021})}\BibitemShut {NoStop}%
\bibitem [{\citenamefont {Afshar}\ and\ \citenamefont {Mazin}(2021)}]{Fe3Ga}%
  \BibitemOpen
  \bibfield  {author} {\bibinfo {author} {\bibfnamefont {M.}~\bibnamefont
  {Afshar}}\ and\ \bibinfo {author} {\bibfnamefont {I.~I.}\ \bibnamefont
  {Mazin}},\ }\bibfield  {title} {\bibinfo {title} {Spin spiral and topological
  hall effect in \ch{Fe3Ga4}.},\ }\href
  {https://doi.org/10.1103/PhysRevB.104.094418} {\bibfield  {journal} {\bibinfo
   {journal} {Phys. Rev. B}\ }\textbf {\bibinfo {volume} {104}},\ \bibinfo
  {pages} {094418} (\bibinfo {year} {2021})}\BibitemShut {NoStop}%
\bibitem [{\citenamefont {Anisimov}\ \emph {et~al.}(1991)\citenamefont
  {Anisimov}, \citenamefont {Zaanen},\ and\ \citenamefont
  {Andersen}}]{hubbard_U}%
  \BibitemOpen
  \bibfield  {author} {\bibinfo {author} {\bibfnamefont {V.~I.}\ \bibnamefont
  {Anisimov}}, \bibinfo {author} {\bibfnamefont {J.}~\bibnamefont {Zaanen}},\
  and\ \bibinfo {author} {\bibfnamefont {O.~K.}\ \bibnamefont {Andersen}},\
  }\bibfield  {title} {\bibinfo {title} {Band theory and mott insulators:
  Hubbard $\mathrm{U}$ instead of stoner $\mathrm{I}$},\ }\href
  {https://doi.org/10.1103/PhysRevB.44.943} {\bibfield  {journal} {\bibinfo
  {journal} {Phys. Rev. B}\ }\textbf {\bibinfo {volume} {44}},\ \bibinfo
  {pages} {943} (\bibinfo {year} {1991})}\BibitemShut {NoStop}%
\bibitem [{\citenamefont {Anisimov}\ \emph {et~al.}(1997)\citenamefont
  {Anisimov}, \citenamefont {Aryasetiawan},\ and\ \citenamefont
  {Lichtenstein}}]{Anisimov_1997}%
  \BibitemOpen
  \bibfield  {author} {\bibinfo {author} {\bibfnamefont {V.~I.}\ \bibnamefont
  {Anisimov}}, \bibinfo {author} {\bibfnamefont {F.}~\bibnamefont
  {Aryasetiawan}},\ and\ \bibinfo {author} {\bibfnamefont {A.~I.}\ \bibnamefont
  {Lichtenstein}},\ }\bibfield  {title} {\bibinfo {title} {First-principles
  calculations of the electronic structure and spectra of strongly correlated
  systems: the $\mathrm{LDA + U}$ method},\ }\href
  {https://doi.org/10.1088/0953-8984/9/4/002} {\bibfield  {journal} {\bibinfo
  {journal} {Journal of Physics: Condensed Matter}\ }\textbf {\bibinfo {volume}
  {9}},\ \bibinfo {pages} {767} (\bibinfo {year} {1997})}\BibitemShut {NoStop}%
\bibitem [{\citenamefont {Galler}\ and\ \citenamefont
  {Pourovskii}(2022)}]{Galler_2022}%
  \BibitemOpen
  \bibfield  {author} {\bibinfo {author} {\bibfnamefont {A.}~\bibnamefont
  {Galler}}\ and\ \bibinfo {author} {\bibfnamefont {L.~V.}\ \bibnamefont
  {Pourovskii}},\ }\bibfield  {title} {\bibinfo {title} {Electronic structure
  of rare-earth mononitrides: quasiatomic excitations and semiconducting
  bands},\ }\href {https://doi.org/10.1088/1367-2630/ac6317} {\bibfield
  {journal} {\bibinfo  {journal} {New Journal of Physics}\ }\textbf {\bibinfo
  {volume} {24}},\ \bibinfo {pages} {043039} (\bibinfo {year}
  {2022})}\BibitemShut {NoStop}%
\end{thebibliography}%

\end{document}



\title{Supplemental Material:\\Topological Hall effect induced by chiral fluctuations in \ch{ErMn6Sn6}}

\author{Kyle~Fruhling$^{\ast\dagger}$}
\affiliation{Department of Physics, Boston College, Chestnut Hill, MA 02467, USA}
\thanks{These authors contributed equally to this work.\\$\dagger$ Corresponding author: fruhling@bc.edu}

\author{Alenna~Streeter$^\ast$}
\affiliation{Department of Physics, Boston College, Chestnut Hill, MA 02467, USA}

\author{Sougata~Mardanya}
\affiliation{Department of Physics and Astrophysics, Howard University, Washington, DC 20059, USA}

\author{Xiaoping~Wang}
\affiliation{Neutron Scattering Division, Oak Ridge National Laboratory, Oak Ridge, TN 37831, USA}

\author{Priya~Baral}
\affiliation{Laboratory for Neutron Scattering and Imaging (LNS), Paul Scherrer Institut, PSI, Villigen, CH-5232, Switzerland}

\author{Oksana~Zaharko}
\affiliation{Laboratory for Neutron Scattering and Imaging (LNS), Paul Scherrer Institut, PSI, Villigen, CH-5232, Switzerland}

\author{Igor~I.~Mazin}
\affiliation{Department of Physics and Astronomy and Quantum Science and Engineering Center, George Mason University, Fairfax, VA 22030, USA}

\author{Sugata~Chowdhury}
\affiliation{Department of Physics and Astrophysics, Howard University, Washington, DC 20059, USA}

\author{William~D.~Ratcliff}
\affiliation{NIST Center for Neutron Research, National Institute of Standards and Technology, Gaithersburg, MD 20899-6100, USA}
\affiliation{Department of Physics and Department of Materials Science and Engineering, University of Maryland, College Park, MD 20742, USA}

\author{Fazel~Tafti}
\affiliation{Department of Physics, Boston College, Chestnut Hill, MA 02467, USA}



\maketitle


\pagebreak

\section{\label{sec:methods} Methods}
\emph{Crystal growth.}
Single crystals of \ch{ErMn6Sn6} were grown using a self-flux technique.
Erbium pieces (99.9\%), manganese granules (99.98\%), and tin pieces (99.999\%) were mixed with the ratio Er:Mn:Sn = 1.25:6:18 and placed in an alumina crucible inside an evacuated quartz tube.  
The excess amount (25\%) of rare-earth was necessary to eliminate impurity phases such as Mn$_3$Sn$_2$ and ErMnSn$_2$.
The quartz tube was heated in a box furnace to 1000\C\ at 3\C/min, held at that temperature for 12h, cooled to 600\C\ at 6\C/h, and centrifuged to remove the excess flux.
Using less amounts of rare-earth or centrifuging at $T<600$\C\ resulted in the formation of impurity phases.

\emph{Characterizations.} Powder x-ray diffraction (PXRD) was performed using a Bruker D8 ECO instrument in the Bragg-Brentano geometry, using a copper source (Cu-K$_{\alpha}$) and a LYNXEYE XE 1D energy dispersive detector.
The FullProf suite was used for the Rietveld analysis~[20] and the VESTA program was used for crystal visualizations~[21].
Magnetization was measured using a Quantum Design MPMS3 with the sample mounted on a low-background quartz holder.
The electrical resistivity and Hall effect were measured in a four-probe configuration using a Quantum Design PPMS Dynacool.

\emph{First-Pinciples Calculations}
The magnetic and electronic structures of \ch{(Y,Er,Tb)Mn6Sn6} we calculated from first-principle by including spin-orbit coupling within the framework of density functional theory (DFT)~[22] using the Vienna {\it ab-initio} simulation package (VASP)~[23, 24]. 
The ground state electronic structure was obtained with the projector augmented-wave pseudopotential, while the electron exchange-correlation effects were implemented by the generalized gradient approximation (GGA)~[25] with Perdew-Burke-Ernzerhof (PBE) parametrization.
The energy cut-off of 350 eV was used for the plane-wave basis set, and the Brillouin zone (BZ) integration was performed with a 9$\times$9$\times$7 $\Gamma$-centered $k-$mesh~[26].
The total energy tolerance criteria are set to $10^{-8}$~eV to satisfy self-consistency. 
We utilized experimental structure parameters and optimized ionic positions until the residual forces on each ion were less than 10$^{-2}$ eV/\AA $~$ and the stress tensors became negligible. 

In our calculations, we considered the inadequacy of DFT methods in respecting the 2nd and 3rd Hund's rules~[4]. 
The problem is that these methods include spurious self-interaction of f-orbitals, which, because of their strong localization, is larger than the spin-orbit and crystal field effects. 
As a result, the occupation state of the f-shell becomes unstable and deviates from the Hund's rules depending on the magnetic structure. 
To circumvent this problem,  we used a patch to the VASP~[23, 24] code, which 
allows nudging DFT+U calculations into a given orbital configuration~[27].

\emph{Neutron Single Crystal Diffraction.}
Zero-field neutron diffraction data for \ch{ErMn6Sn6} (Sample 1) were collected on the time-of-flight single crystal diffractometer, TOPAZ, at the Spallation Neutron Source, Oak Ridge National Laboratory~[28]. 
A plate-shaped crystal measuring 0.49 x 3.16 x 3.25 mm was securely attached to a custom-built aluminum pin with SuperGlue and then mounted on the TOPAZ ambient goniometer for data collection at 200 K, with its temperature controlled by an LN2 Cobra Cryostream. Data collection at 5 K used the TOPAZ cryogenic goniometer, with sample temperature controlled by a Cryomech P415 pulse tube cryocooler. 
Crystal orientations were optimized with the CrystalPlan program~[29] (19 at 200 K and 20 at 5 K). 
Measurement for each orientation used approximately 1 hour of neutron beamtime, with 5 Coulombs of proton charge for SNS beam power at 1.4 MW. A multiresolution machine learning algorithm was used for peak integration. 
Data normalization, including neutron time-of-flight spectrum, Lorentz, and detector efficiency corrections, followed previously reported procedures~[30]. 
A spherical absorption correction with $\mu = 0.02155 + 0.05705\lambda$ mm$^{-1}$ was applied. The reduced data were saved in SHELX HKLF2 format, with the wavelength for each reflection recorded separately and not merged. Magnetic structure solution and refinement were performed using the JANA2020 program~[31]. 
The crystal structures at 200 K and 40 K from X-ray diffraction, both of $P6/mmm$ parent space group symmetry, served as the starting models for magnetic structure refinements. 

Neutron diffraction data under a magnetic field were measured on the thermal-neutron single crystal diffractometer ZEBRA at SINQ of Paul Scherrer Insitut (Villigen, Switzerland). 
The instrument was operated in the normal beam geometry. 
An incoming neutron wavelength of 1.383 \AA\ (Ge monochromator) was used. 
A crystal of dimensions 3.25 mm$\times$2.83 mm$\times$1 mm was mounted in  vertical 6 T or horizontal 7 T magnets with the direct space $a$ crystal axis vertical. 
In addition to temperature and magnetic field scans, data sets in magnetic fields applied vertically were collected. 
28 and 23 magnetic reflections were measured at $T$= 200 K and $H$=1.5 T and $H$=3.5 T, respectively, and were used for refinements of magnetic models using FULLPROF and JANA.

\section{\label{sec:magnetization}Magnetization Data}
\begin{figure*}[ht]
 \centering
 \includegraphics[width=\textwidth]{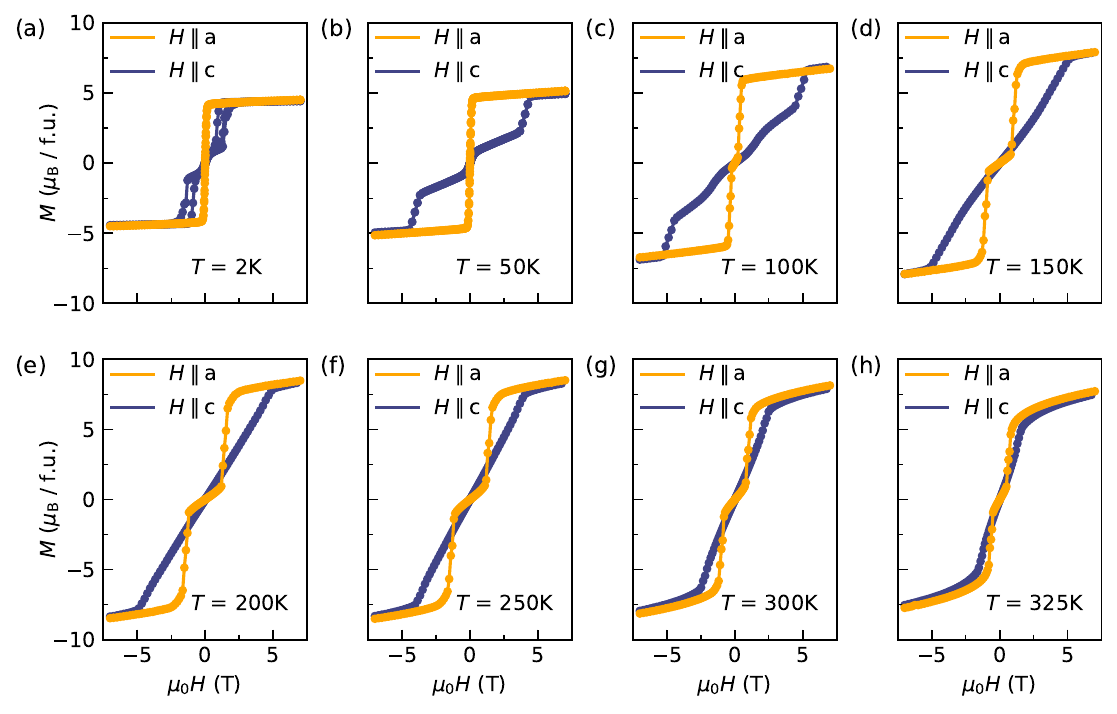}
 \caption{\label{fig:MH} 
 (a-h) Magnetization curves in both field directions $H\|a$ (orange) and $H\|c$ (blue) are shown at several temperatures.
 }
\end{figure*}
\begin{figure*}[ht]
 \centering
 \includegraphics[width=\textwidth]{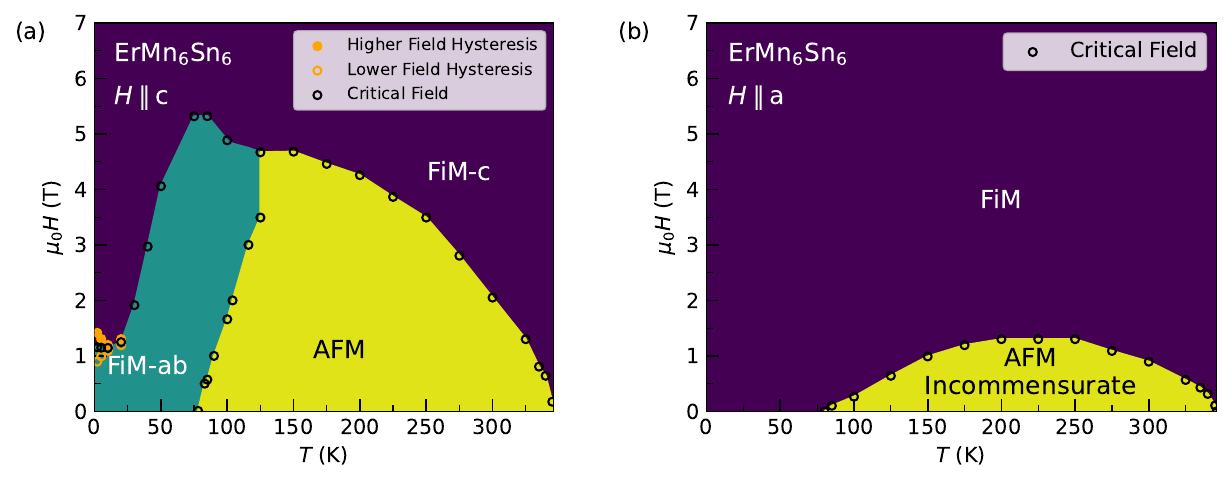}
 \caption{\label{fig:PD_ab_c} 
 (a) Magnetic phase diagram with $H\|c$. (b) Magnetic phase diagram with $H\|a$.
 }
\end{figure*}

Fig.~\ref{fig:MH} shows magnetization curves at different temperatures for both $H\|a$ (orange) and $H\|c$ (blue) from which we constructed the phase diagrams in Fig.~\ref{fig:PD_ab_c}.
At low temperatures with $H\|c$, there is notable hysteresis. At these low temperatures, the critical field is defined as the average of the critical field with field increasing and of the critical field with field decreasing.
Our discussion in the main text was focused on the phase diagram for $H\|a$ shown in Fig.~\ref{fig:PD_ab_c}b and reproduced in the main text (Fig.~1f). 

\section{\label{sec:zebra}Neutron Diffraction Data from the Zebra Experiment}
\begin{figure*}[ht]
 \centering
 \includegraphics[width=\textwidth]{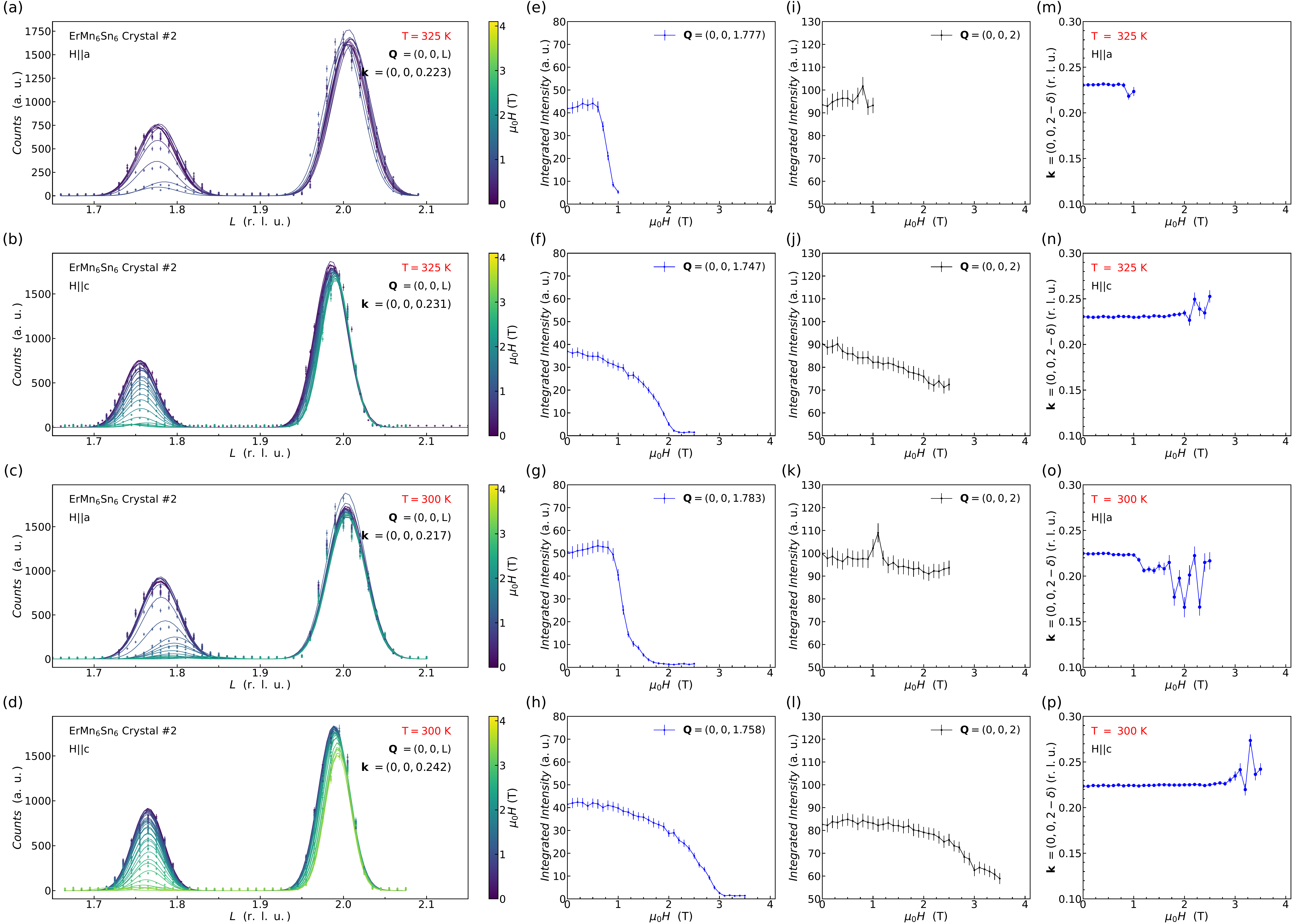}
 \caption{\label{fig:N1} 
 (a-d) $\mathbf{Q}$-scans obtained around the (002) reflection to reveal the magnetic satellite peak at 325 and 300 K with $H\|a$ and $H\|c$ as indicated on each panel. Different colors show different field strengths. 
 (e-h) Integrated intensity of the satellite peak as a function of field strength corresponding to panels a-d, respectively.
 (i-l) Integrated intensity of the (002) peak as a function of field strength corresponding to panels a-d, respectively.
 (m-p) Incommensurate propagation wavevector as a function of field strength for panels a-d, respectively.
 }
\end{figure*}
Here we show the complete set of neutron diffraction data used to construct the phase diagrams with $H\|a$ and $H\|c$.
Figs.~\ref{fig:N1}a and \ref{fig:N1}b show the (002) reflection and magnetic satellite peaks at $T=325$~K with magnetic fields along the $a$ and $c$ axis directions, respectively.
Note that only one of the satellites, i.e. the one at $(0,0,L-\mathbf{k})$, is shown, because measuring both satellites was too time consuming within the beamtime period we had for measurements.
In both directions, increasing the magnetic field suppresses the satellite peaks.
This suppression is completed at 1 T for $H\|a$ and 2 T for $H\|c$. 
Each peak is fit with a Gaussian peak shape. 
Similar data are shown in Figs.~\ref{fig:N1}c ($H\|a$) and \ref{fig:N1}d ($H\|c$) at $T=300$ K.

Panels e and f show the integrated intensity of the incommensurate magnetic peaks (satellite peaks) at $T=325$~K in panels a and b as a function of field strength.
These satellite peaks are observed at $\mathbf{Q}=(0,0,1.777)$ and $\mathbf{Q}=(0,0,1.747)$ as stated on Figs.~\ref{fig:N1}e and \ref{fig:N1}f. 
For $H\|a$ (Fig.~\ref{fig:N1}e), the suppression takes on a sigmoidal shape with a sharp decrease in intensity before 1 T. 
For $H\|c$ (Fig.~\ref{fig:N1}f), a more gradual power law behavior is observed with complete suppression at 2 T. 
Similar behaviors are observed at $T=300$ K in Figs.~\ref{fig:N1}g and \ref{fig:N1}h, with a sigmoidal suppression of the satellite peaks for $H\|a$ and a power-law suppression for $H\|c$.

Panels i and j show the integrated intensity of the (002) nuclear peak as a function of field strength at $T=325$ K.  
For $H\|a$ (Fig.~\ref{fig:N1}i), the nuclear intensity remains relatively unchanged in field. 
We also note that Crystal \#2 shows a small intensity peak at the critical field, a feature which does not occur in Crystal \#1. 
We expect the intensity to change in the orthogonal directions as the FM component is induced along c, but we could not measure these directions due to geometric limitations of the experimental setup.
For $H\|c$ (Fig.~\ref{fig:N1}j), the nuclear intensity decreases gradually with increasing field. 
This may be an indication that there is magnetic intensity on the nuclear peaks from an additional $k = (0,0,0)$ order at zero-field, which is re-distributed to in-plane reflections when the magnetic field polarizes spins out of plane towards the $c$-axis.  
Similar observations are made at $T=300$~K in Figs.~\ref{fig:N1}k and \ref{fig:N1}l.

The magnetic propagation vector is given by the difference $\delta$ between the two peak positions and is displayed as a function of field at $T=325$ K in panels m and n. 
For $H\|a$ (Fig.~\ref{fig:N1}m), there is a subtle shift in the satellite peak position at the transition field, illustrated by the dip in $\delta$.  
The shift, which is more noticeable at lower temperatures, indicates an intermediate magnetic phase between the AFM and FIM orders in the phase diagram of Fig.~\ref{fig:PD_ab_c}b. 
There is no shift in the propagation vector for $H\|c$ (Figure \ref{fig:N1}n).

\begin{figure*}[ht]
 \centering
 \includegraphics[width=\textwidth]{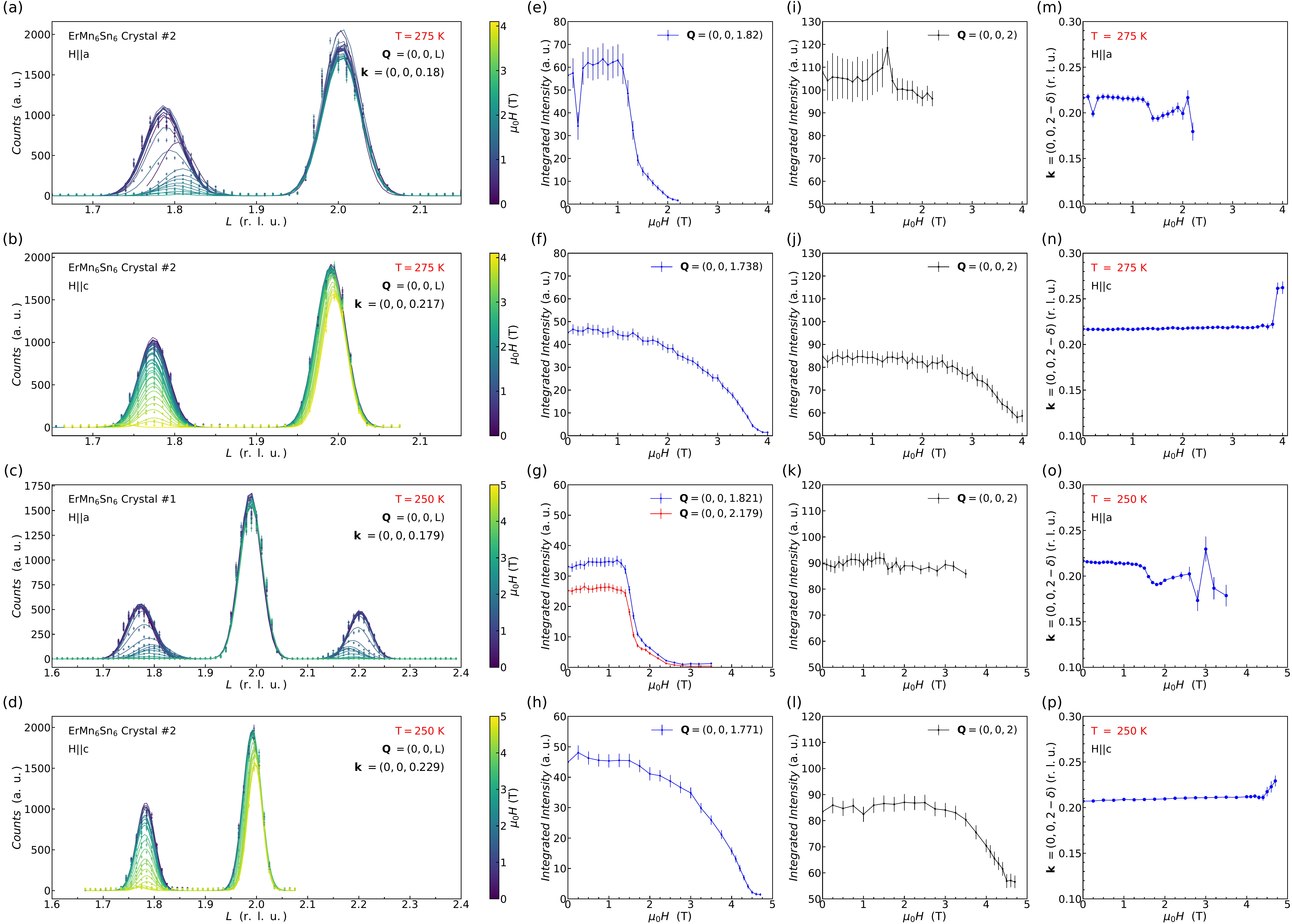}
 \caption{\label{fig:N2} 
 (a-d) $\mathbf{Q}$-scans obtained around the (002) reflection to reveal the magnetic satellite peak at 275 and 250 K with $H\|a$ and $H\|c$ as indicated on each panel. Different colors show different field strengths. 
 (e-h) Integrated intensity of the satellite peak as a function of field strength corresponding to panels a-d, respectively.
 (i-l) Integrated intensity of the (002) peak as a function of field strength corresponding to panels a-d, respectively.
 (m-p) Incommensurate propagation wavevector as a function of field strength for panels a-d, respectively.
 }
\end{figure*}

The same features at 325 and 300 K are also present at 275 and 250 K as shown in Fig.~\ref{fig:N2}.
Several systematic trends in Figs.~\ref{fig:N1} and \ref{fig:N2} can be summarized as follows.
(i) The scans shown in Figs.~\ref{fig:N1}a-d and \ref{fig:N2}a-d include both the (0,0,2) nuclear Bragg peak and a smaller satellite peak at an incommensurate $\mathbf{Q}=(0,0,2-\delta)$.
This general structure is best observed in Fig.~\ref{fig:N2}c with the nuclear (0,0,2) peak at the center and two satellite peaks on either side of it.
That scan was obtained for Crystal \#1 at 250 K and with $H\|a$.
Due to experimental time constraint, we only measured the satellite peak at the smaller $\mathbf{Q}$-vector at all other temperatures and field directions.
(ii) The suppression of the satellite peak intensity with field always follows a sigmoidal behavior for $H\|a$ (Panels e and g) and a power law behavior for $H\|c$ (Panels f and h).
(iii) The nuclear (0,0,2) peak does not show any field dependence when $H\|a$ (Panels i and k), but it shows a mild suppression when $H\|c$ (Panels j and l).
From here, we conclude that the magnetic structure is in-plane for $H\|a$, but when $H\|c$ the spins are canting out of plane and the magnetic structure acquires a FM component.

\section{\label{sec:zebra}Neutron Diffraction Data from the TOPAZ Experiment}
We used the TOPAZ single-crystal diffractometer to refine the magnetic structure of \ch{ErMn6Sn6} in both FIM and AFM phases (see the phase diagram of Fig.~\ref{fig:PD_ab_c}b) in zero-field.
Prior to the refinement analysis of neutron data, we determined the crystal structure at 40~K and 200~K from X-ray diffraction, and found it to be in the $P6/mmm$ space group symmetry at both low and high temperatures.
We used this structural model as the starting point for the magnetic refinements of neutron single crystal diffraction data measured at 5~K and 200~K in Tables~\ref{tab:topaz_refine_5K} and \ref{tab:topaz_refine_200K}. 

\begin{figure*}[ht]
 \centering
 \includegraphics[width=\textwidth]{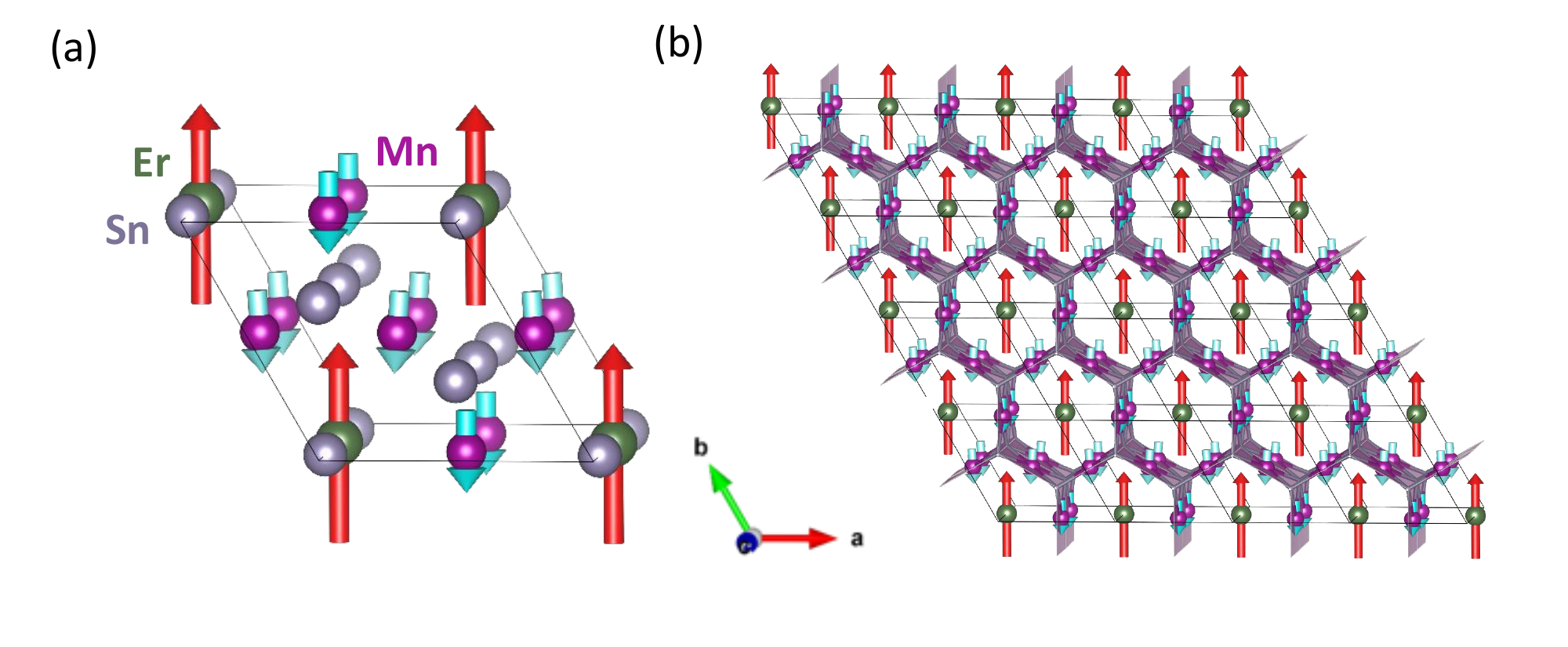}
 \caption{\label{fig:5K_Mag_Structure} 
 (a) Ferrimagnetic structure with all atoms in the space group $Cmm'm'$.
 (b) Ferrimagnetic structure with Sn atoms omitted for clarity. The Mn moments in the Kagome lattice comprising (\ch{Mn3Sn2})$_2$ subunits interact antiferromagnetically with Er atoms in the channels. 
 }
\end{figure*}

First, we discuss the magnetic structure at 5~K, i.e. the magnetic ground state of \ch{ErMn6Sn6}.
As listed in Table~\ref{tab:topaz_refine_5K}, we found fits of the data in two magnetic space groups, namely a $C2'/m'$ space group and a $Cmm'm'$.
Since the latter has a higher symmetry than the former, we identify $Cmm’m’$ as the magnetic ground state of the material.
Individual components of the magnetic moment for different Er and Mn sites are listed in Table~\ref{tab:topaz_moments_5K} for this structure, which is referred to as the FIM order in the main manuscript.
Figure~\ref{fig:5K_Mag_Structure} illustrates the FIM arrangement of magnetic moments at 5~K.

\begin{figure*}[ht]
 \centering
 \includegraphics[width=\textwidth]{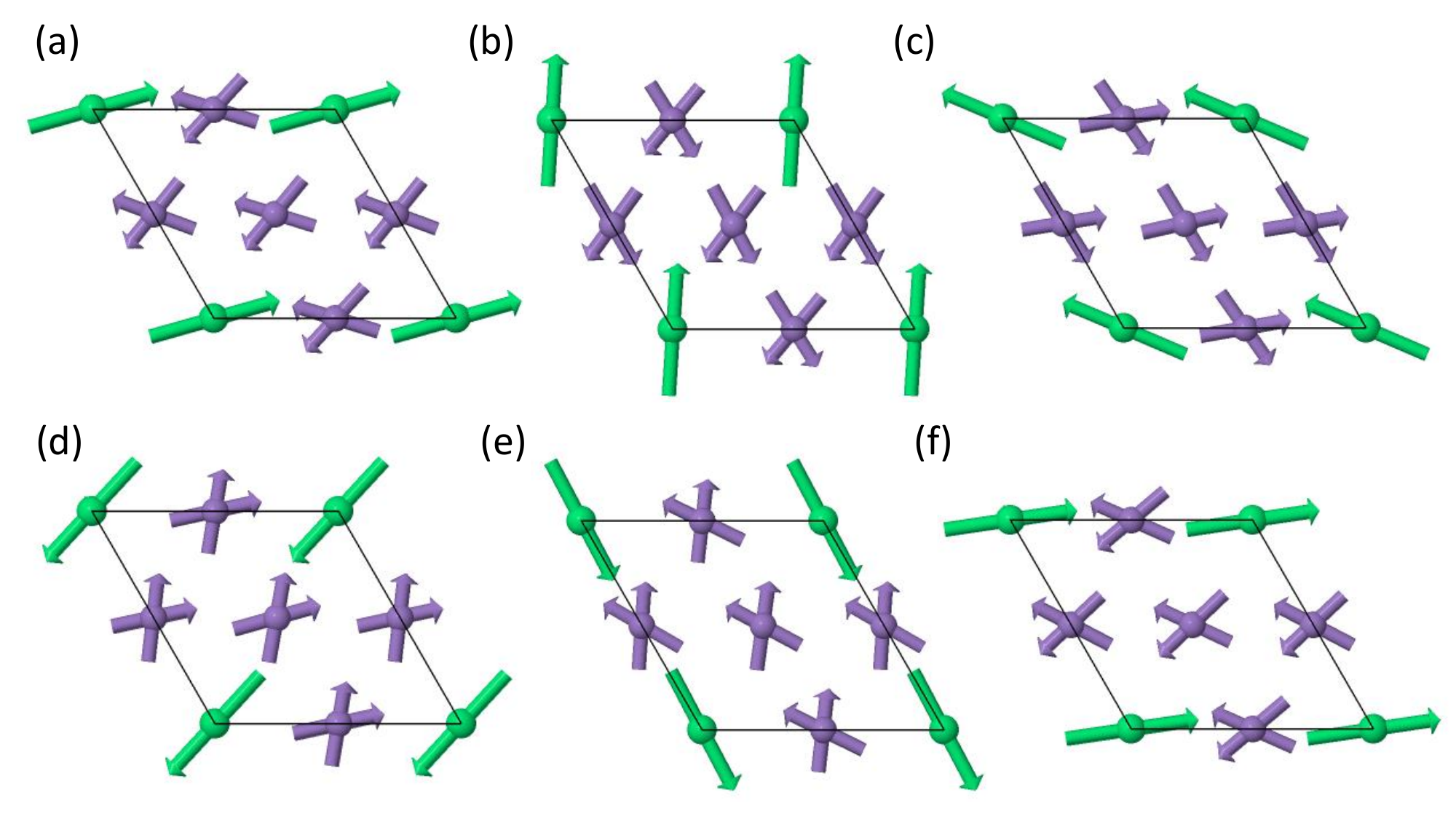}
 \caption{\label{fig:SPIRAL} 
 Magnetic structure model for \ch{ErMn6Sn6} solved in \textit{P622.1’(00g)h00s} is shown in panels (a--f), where each panel shows one Mn-Er-Mn sandwich layer. The figure shows how the spiral evolves and repeats itself after 5 Mn-Er-Mn layers.
 }
\end{figure*}

Next, we discuss the magnetic structure at 200~K.
We found two magnetic models that could fit the neutron diffraction data at 200~K (Table~\ref{tab:topaz_refine_200K}).
The first model in the magnetic space groups $P622.1’(00g)h00s$ has a spiral AFM order as shown in Fig.~\ref{fig:SPIRAL}.
Individual components of the magnetic moment for different Er and Mn sites are listed in Table~\ref{tab:topaz_moments_200K} for this structure, which is referred to as the spiral AFM order in the main manuscript.
The second model in the magnetic space group $Cmmm.1’(00g)s00s$ also has a spiral for the Mn moments.  However, the Er moments have a collinear arrangement of spins with amplitude-modulated moments.
We rule out this magnetic structure based on two reasons: first, it implies that the Mn and Er sublattices are independent from each other which contradicts the DFT results; second, this is a lower symmetry solution compared to the double-spiral structure.
As such, we only show the double spiral AFM order associated with $P622.1’(00g)h00s$ in the main manuscript. 

\begin{table*}[htbp]
  \centering
   \caption{\label{tab:topaz_refine_5K}Crystal data and magnetic structure refinement details for \ch{ErMn6Sn6} at 5~K and zero-field.}
  \begin{tabular}{lcc}
    \hline
    Chemical formula & \ch{ErMn6Sn6} \\
    $M_r$ & 1209.1 & \\
    Parent crystal system, space group & Hexagonal, \textit{P6/mmm} & \\
    Temperature (K) & 5 & \\
    \textit{a}, \textit{c} (\AA) & 5.4934 (4), 8.9700 (9) & \\
    \textit{V} (\AA\textsuperscript{3}) & 234.43 (3) \\
    \textit{Z} & 1 & \\
    Propagation vector & \textit{\textbf{k}} = (0, 0, 0) &\\
    Magnetic (super)space group No. & 65.486 & 12.62\\
    BNS Magnetic (super)space group & \textit{Cmm’m’} & \textit{C2'/m'}\\
    Transformation to a standard setting & \textbf{a}+2\textbf{b}, -\textbf{a}, \textbf{c} | (0,0,0) & -\textbf{a}-2\textbf{b}, \textbf{a}, \textbf{c} | (0,0,0)\\
    \textit{a, b, c} (\AA) & 9.5148, 5.4934, 8.9700 \\
    Active Irreps & \\
    Radiation type & TOF Neutron, $\lambda$ = 0.5 – 3.5 \AA & \\
    $\mu$ (mm\textsuperscript{-1}) & 0.02155 + 0.05705$\lambda$ & \\
    Crystal size (mm) & 3.29 $\times$ 3.16 $\times$ 0.49 & \\
    Diffractometer & SNS BL-12 TOPAZ & \\
    Tmin, Tmax & 0.625, 0.909 & \\
    No. of measured, independent, and & \\ 
    observed [I $>$ 3$\sigma$(I)] reflections & 5814, 5814, 5693 & 5814, 5814, 5694\\
    R[F\textsuperscript{2} $>$ 2$\sigma$(F\textsuperscript{2})], wR(F\textsuperscript{2}), S & 0.052, 0.132, 1.12 & 0.050, 0.125, 1.13\\
    No. of reflections & 5693 & 5694\\
    No. of parameters & 32 & 44\\
    $\Delta\rho_{\text{max}}$, $\Delta\rho_{\text{min}}$ (e \AA\textsuperscript{-3}) & 0.36, $-$0.31 & 0.37, $-$0.46\\
    \hline
  \end{tabular}
\end{table*}

\begin{table*}[ht]
\centering
\caption{\label{tab:topaz_moments_5K} Magnetic parameters of \ch{ErMn6Sn6} at 5 K and zero-field, \textbf{\textit{k}} = (0, 0, 0) in the space group $Cmm'm'$.}
Magnetic Moments \\ 
\begin{tabular}{lcccc}
\hline
Atom   & Mx         & My         & Mz  &  |M| \\ \hline
Er1    &  4.6178(2) & 9.2355(3)  & 0.00   &  \textbf{7.9982(4)} \\
Mn1\_1 & -1.340(2)  & -2.680(4)  & 0.00   &  \textbf{2.321(5)} \\
Mn1\_2 & -1.189(4)  & -2.690(2)  & 0.00   &  \textbf{2.335(5)} \\ \hline\\
\end{tabular}

The sum of magnetic moments over the whole cell\\
\begin{tabular}{lccc}
\hline
Atom   & Mx         & My         & Mz \\ \hline
Er1    &  4.6178(2) & 9.2355(3)  & 0.00  \\
Mn1\_1 & -2.680(3)  & -5.360(3)  & 0.00  \\
Mn1\_2 & -5.379(6)  & -10.759(3) & 0.00  \\ \hline
Sum    & \textbf{-3.441(7)} & \textbf{-6.883(4)}  & \textbf{0.00} \\ \hline\\
\end{tabular}
\end{table*}

\begin{table*}[htbp]
  \centering
  \caption{\label{tab:topaz_refine_200K}Refinement details for the magnetic structure of \ch{ErMn6Sn6} at 200 K and zero-field.}
  \begin{tabular}{lcc}
    \hline
    Chemical formula & \ch{ErMn6Sn6} &\\
    $M_r$ & 1209.1 &\\
    Parent crystal system, space group & Hexagonal, \textit{P6/mmm} &\\
    Temperature (K) & 200 &\\
    \textit{a}, \textit{c} (\AA) & 5.5028 (4), 8.9726 (9) &\\
    \textit{V} (\AA\textsuperscript{3}) & 235.30 (3) &\\
    \textit{Z} & 1  &\\
    Propagation vector & \textit{\textbf{k}} = (0, 0, 0.1959) &\\
    Magnetic (super)space group No. & 177.1.24.2.m150.2 & 65.1.13.2.m482.2\\
    BNS Magnetic (super)space group & \textit{P622.1'(0,0,g)h00s} & \textit{Cmmm.1'(0,0,g)s00s} \\
    Transformation to a standard setting & \textbf{a}, \textbf{b}, \textbf{c} | (0,0,0,1/6) & \textbf{a}, \textbf{a}+2\textbf{b}, \textbf{c} | (0,0,0,0)\\
    \textit{a, b, c} (\AA) & 5.5028, 5.5028, 8.9726 & 5.5028, 9.5311, 8.9726\\
    Active Irreps & & \\
    Radiation type & Neutron, $\lambda = 0.4004-3.4951$ Å &\\
    $\mu$ (mm\textsuperscript{-1}) & 0.02155 + 0.05705$\lambda$ &\\
    Crystal size (mm) & 3.29 $\times$ 3.16 $\times$ 0.49 &\\
    Diffractometer & SNS BL-12 TOPAZ &\\
    Tmin, Tmax & 0.624, 0.909 &\\
    No. of measured, independent, and &\\
    observed [I $>$ 3$\sigma$(I)] reflections & 5527, 5527, 5233 & 5527, 5527, 5174 \\
    R[F\textsuperscript{2} $>$ 2$\sigma$(F\textsuperscript{2})], wR(F\textsuperscript{2}), S & 0.045, 0.136, 1.59 & 0.043, 0.122, 1.49\\
    No. of reflections & 5233 & 5174\\
    No. of parameters & 25 & 34\\
    $\Delta\rho_{\text{max}}$, $\Delta\rho_{\text{min}}$ (e Å\textsuperscript{-3}) & 0.23, $-$0.33 & 0.30, $-$0.29\\
    \hline
  \end{tabular}
\end{table*}

\begin{table*}[htbp]
  \centering  \caption{\label{tab:topaz_moments_200K}Magnetic parameters of \ch{ErMn6Sn6} at 200 K and zero-field, \textbf{\textit{k}} = (0, 0, 0.1957) in the magnetic superspace group $P622.1'(0,0,g)h00s$.}
  Magnetic moments \\
  \begin{tabular}{lcccccc}
    \hline
    Atom & wave & Mx & My & Mz & |M| \\
    \hline
    Er1 & sin1 & 0.00 & 0.00 & 0.00 & \textbf{0.00} \\
        & cos1 & 2.3604(3) & 4.7209(6) & 0.00 & \textbf{4.0884(7)} \\
    Mn1 & sin1 & -0.2721(10) &    -0.544(2) & 0.00 & \textbf{0.471(2)} \\
        & cos1 & -1.5444(10) &    -3.089(2) & 0.00 & \textbf{2.675(2)} \\
    Mn2 & sin1 & -0.7018(18) &    -1.0566(11) & 0.00 & \textbf{0.931(2)} \\
        & cos1 &  -1.1877(14) &    -2.9373(11)  & 0.00 & \textbf{2.5592(18)} \\
    \hline \\
  \end{tabular}
\end{table*}

\section{\label{sec:the}Topological Hall Effect (THE)}
\begin{figure*}[ht]
 \centering
 \includegraphics[width=\textwidth]{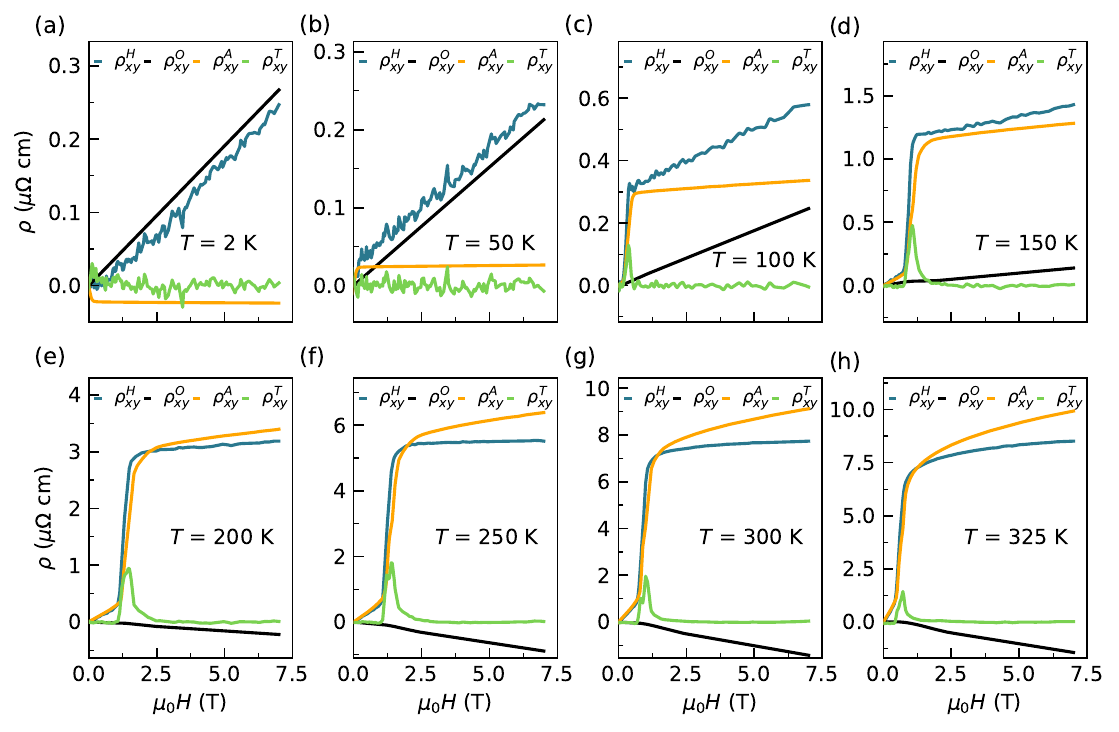}
 \caption{\label{fig:RH} 
 (a-h) The analysis of the THE at different temperatures as described in the main text.
}
\end{figure*}

The color scale in Fig.~3b of the main text represents the size of the THE determined by the analysis shown in Fig.~\ref{fig:RH}.
At each temperature, we extract the topological Hall resistivity ($\rho^T_{xy}$) by subtracting the anomalous and ordinary contributions from the total Hall resistivity, as explained in the main text. 
To create the color map, this analysis was carried out at several temperatures, and then a linear interpolation was applied between THE magnitudes extracted at different temperatures.

\begin{figure*}[ht]
 \centering
 \includegraphics[width=\textwidth]{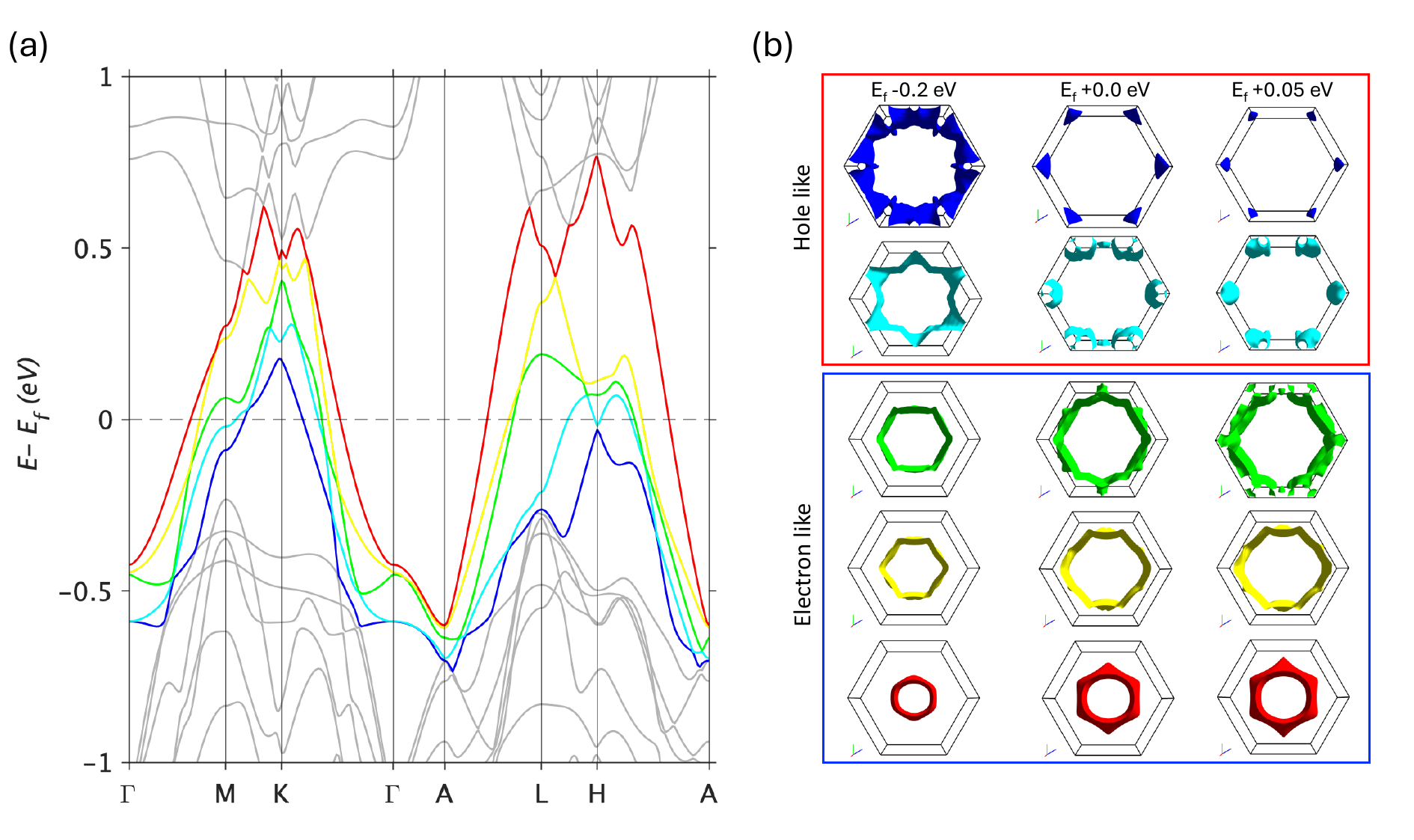}
 \caption{\label{fig:FIM_Surface_Down} 
 (a) Electronic band structure of ferrimagnetic \ch{ErMn6Sn6} with the spin-down component along the high symmetry line of the Brillouin Zone. The bands crossing the Fermi energy are highlighted by different colors. (b) Change in the Fermi surface with a change in chemical potential for each of the bands highlighted in (a). 
 }
\end{figure*}

\begin{figure*}[ht]
 \centering
 \includegraphics[width=\textwidth]{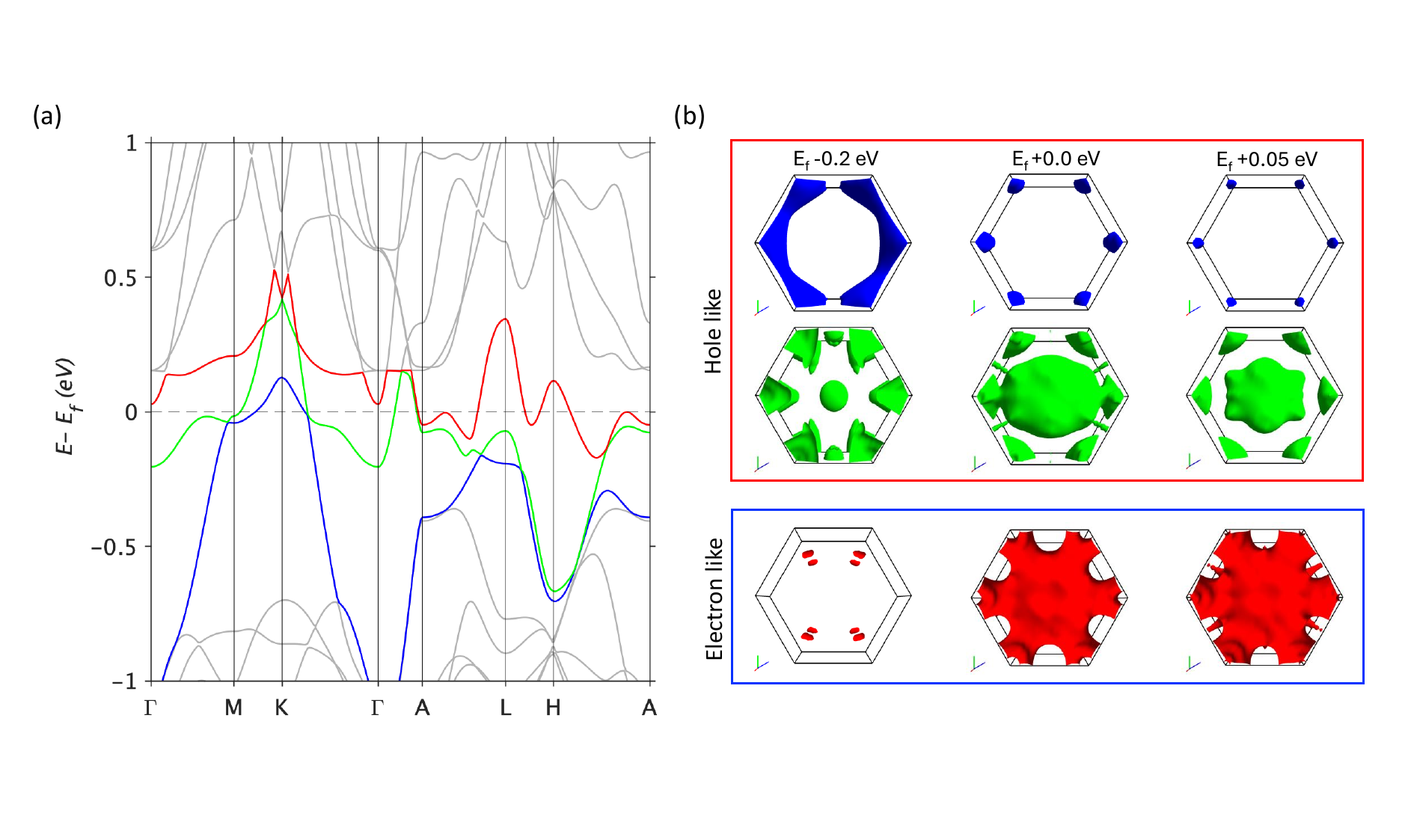}
 \caption{\label{fig:FIM_Surface_Up} 
 (a) Electronic band structure of ferrimagnetic \ch{ErMn6Sn6} with the spin-up component along the high symmetry line of the Brillouin Zone. The bands crossing the Fermi energy are highlighted by different colors. (b) Change in the Fermi surface with a change in chemical potential for each of the bands highlighted in (a). The existence of hole-like Fermi surfaces is robust regardless of if the FIM state is analyzed in a spin-up or spin-down state.
 }
\end{figure*}

\begin{figure*}[ht]
 \centering
 \includegraphics[width=\textwidth]{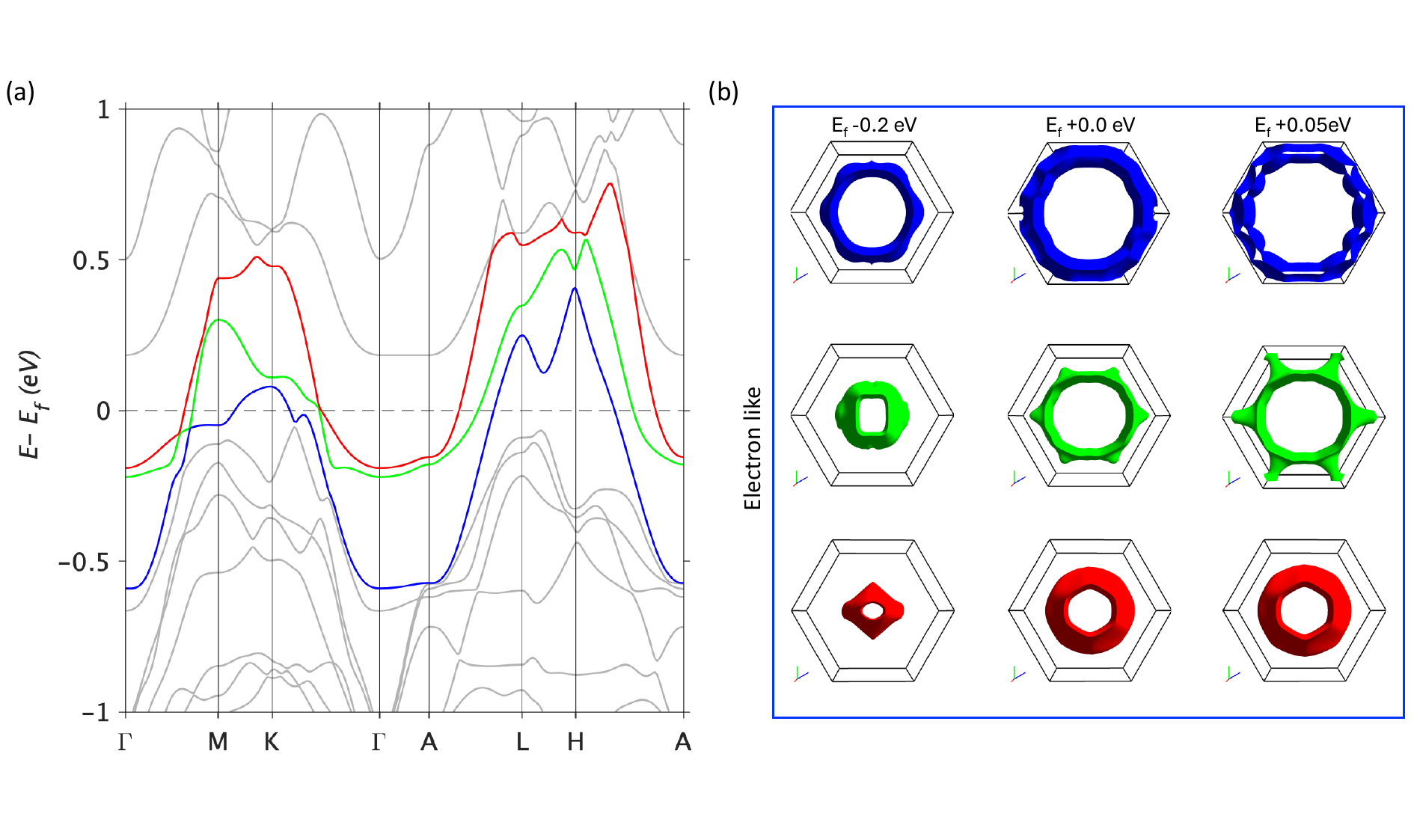}
 \caption{\label{fig:AFM_Surface} 
 (a) Electronic band structure of antiferromagnetic \ch{ErMn6Sn6} along the high symmetry line of the Brillouin Zone. The bands crossing the Fermi energy are highlighted by different colors. (b) Change in the Fermi surface with a change in chemical potential for each of the bands highlighted in (a).
 }
\end{figure*}

\begin{figure*}[ht]
 \centering
 \includegraphics[width=\textwidth]{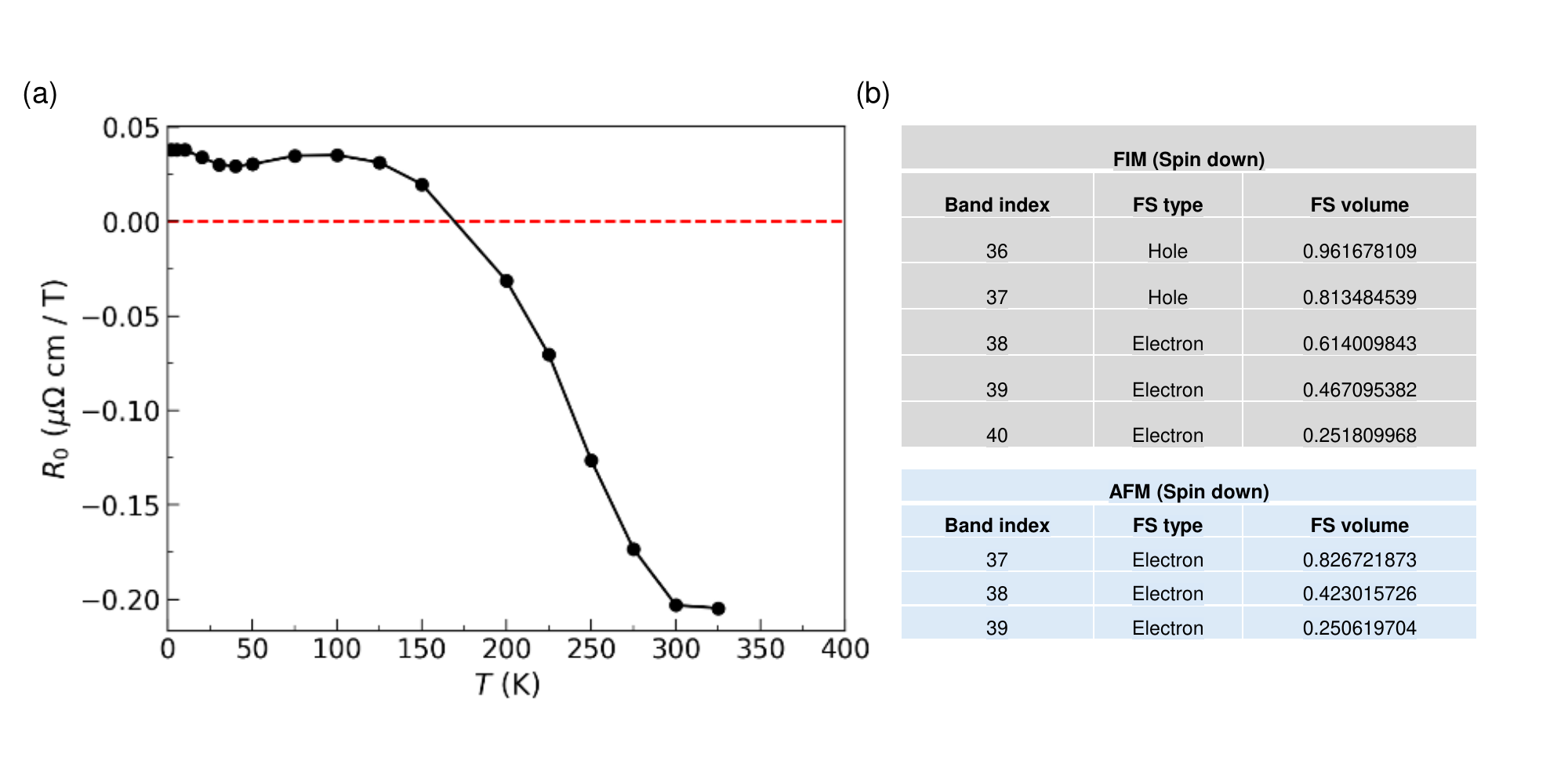}
 \caption{\label{fig:R0} 
 (a) Evolution of the ordinary Hall coefficient with temperature. (b) Fermi energy crossing bands in spin-down FIM and AFM states. Note that the FIM state has both hole like and electron like bands while the AFM state has purely electron like bands.
 }
\end{figure*}

As the temperature increases there is a change in the ordinary Hall resistivity from a positive slope to a negative slope (Fig.~\ref{fig:RH}). This can be explained by the Fermi surface evolving from a hole-dominated one to an electron one. The change in the nature of the Fermi Surface is shown in Figs.~\ref{fig:FIM_Surface_Down}-\ref{fig:AFM_Surface}. The AFM state was approximated by removing the Er moment which is much smaller at higher temperatures. Because, DFT calculations are ground-state calculations the temperature evolution of the high field FIM state cannot be shown. However, the Fermi surface is shown by Figs.~\ref{fig:FIM_Surface_Down}-\ref{fig:AFM_Surface} to move towards an electron like state as the Er moment decreases. That this change is reasonable is supported by the smooth evolution of the ordinary Hall coefficient shown in Fig.~\ref{fig:R0}a. In the FIM ground state there are both electron like and hole like bands contributing to the Fermi surface while in the higher temperature AFM state the bands are purely electron like as summarized in Fig.~\ref{fig:R0}b.

\begin{figure*}[ht]
 \centering
 \includegraphics[width=\textwidth]{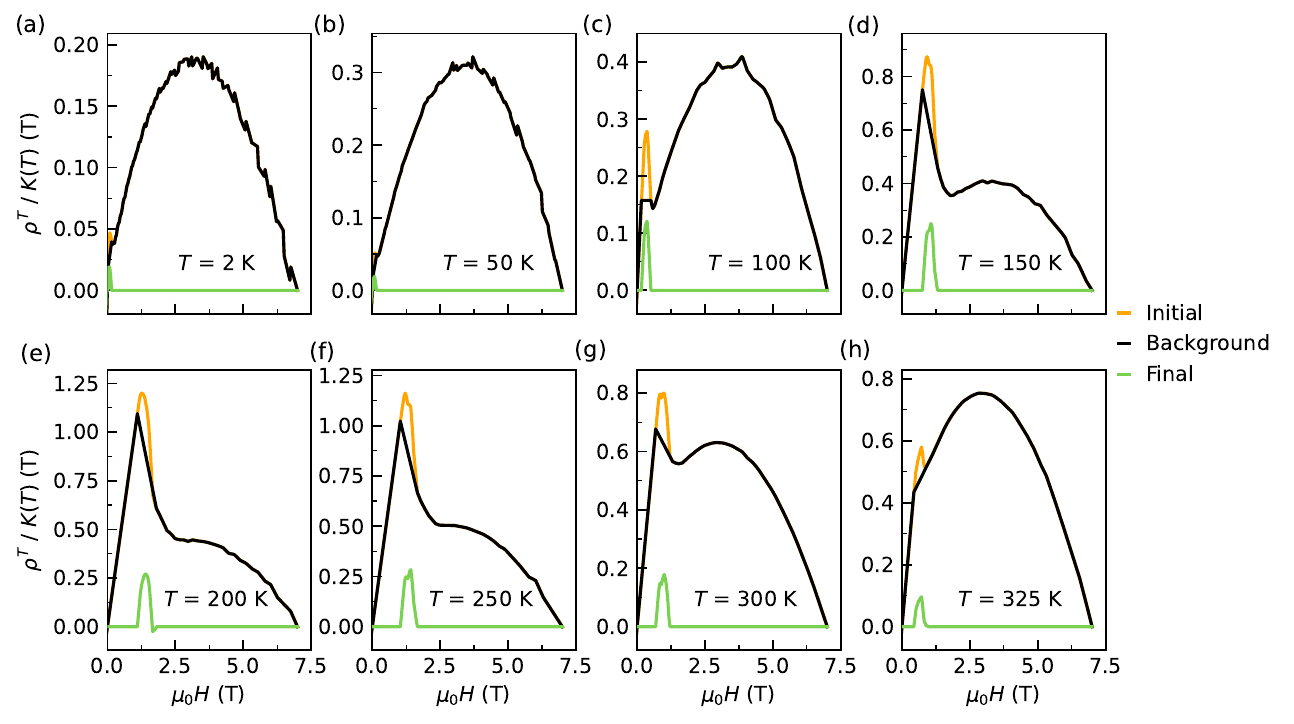}
 \caption{\label{fig:RH3} 
 (a-h) The analysis of the THE using Eq. 1 in the main text.
 }
\end{figure*}

\begin{figure*}[ht]
 \centering
 \includegraphics[width=\textwidth]{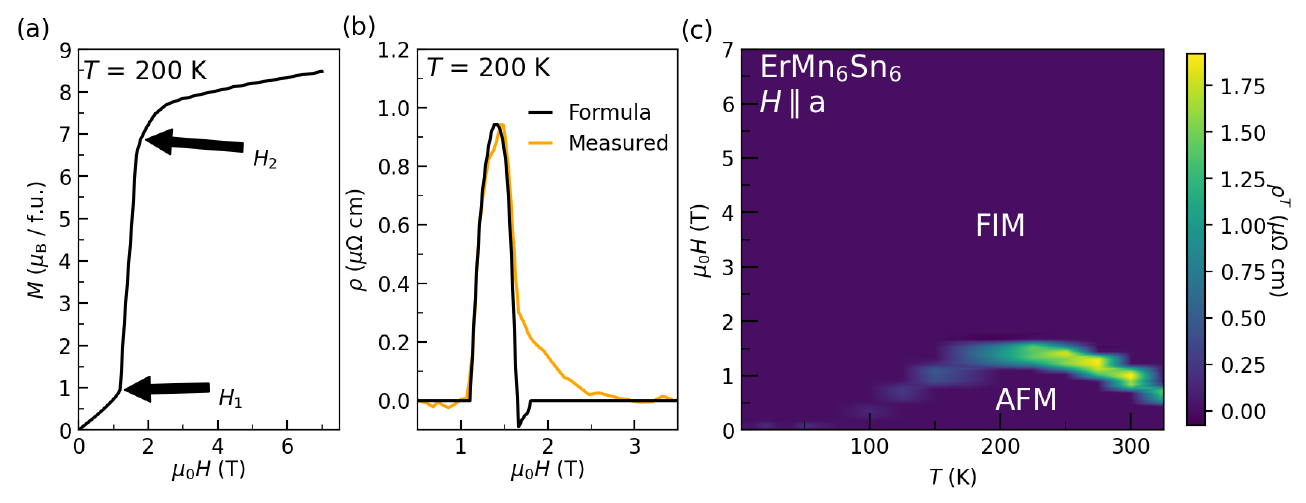}
 \caption{\label{fig:H1H2} 
 (a) Magnetization of \ch{ErMn6Sn6} at 200K with fields $H_1$ and $H_2$ marking the limits of the application of Eq. 1. 
 (b) Comparison of THE extracted from resistivity data and from application of Eq. 1. 
 (c) Phase Diagram constructed using the THE extracted using Eq.~1.
 }
\end{figure*}

The main text also discusses the expected form of the THE given by $\rho^{T}=\kappa\, M_c^2TH$ for a transverse conical spiral (TCS) structure.
Due to the complex temperature dependence of the moment, this cannot be applied directly to all temperatures with some constant $\kappa$, but can be applied at each temperature to compare the location of the peak and width of the measured THE to theory.
The factor $M_c^2$ in the \ch{YMn6Sn6} case can be expressed as $1-\frac{M^2}{M_s^2}$ where $M_s$ is the saturation magnetization. 
In Fig.~\ref{fig:RH3} we show the overall curve given by $\rho^{T}/K(T)=(1-M^2/M_s^2)H$, where $K(T)$ is some function representing the temperature dependence, at a few different temperatures. 
Because this equation is only valid in the TCS phase, we do a linear interpolation between the boundary points of this region (critical fields $H_1$ and $H_2$) to determine a background which is subtracted to give the expected THE.
The critical fields $H_1$ and $H_2$ are defined in Fig.~\ref{fig:H1H2}a as follows:
$H_1$ is the field at which a sudden increase in resistivity is observed (it is marked by a maximum in the second derivative of the resistivity as a function of field).
$H_2$ is marked by a deviation of 10\% from the high field linear fit to the $M(H)$ data.
In Fig.~\ref{fig:H1H2}b, a comparison is made between the scaled theoretical THE and measured THE to show their agreement with respect to location of the peak and width of the peak at the half maximum. 
Notably, the measured THE decreases slower at higher fields instead of forming a sharp peak as in the predicted THE, this is likely due to the more complicated behavior of $M_c^2$ due to the presence of the magnetic Er moment compared to the simpler Y case. 
Finally, we construct another phase diagram of the THE based on the predicted THE values~\ref{fig:H1H2}c which is comparable to the phase diagram constructed by direct measurement.

